\documentclass[12pt,tightenlines,eqsecnum,floats,aps,amsmath,amssymb,tightenlines,nofootinbib,prd,shownopacs,notitlepage,superscriptaddress]{revtex4-1}

\usepackage{graphicx,verbatim}
\usepackage{amsmath}
\usepackage{amsfonts}
\usepackage{amssymb}
\usepackage[colorinlistoftodos]{todonotes}
\usepackage{epstopdf}

\usepackage{bm}
\usepackage[applemac]{inputenc}
\usepackage[spanish,english]{babel}
\usepackage{amsmath,amssymb,amsfonts,cancel}

\usepackage{color}
\usepackage{soul}
\usepackage{hyperref}
\usepackage{slashed}

\allowdisplaybreaks

\newcommand*{\Scale}[2][4]{\scalebox{#1}{$#2$}}
\newcommand{\be}{\begin{equation}}
\newcommand{\ee}{\end{equation}}
\newcommand{\bea}{\begin{eqnarray}}
\newcommand{\eea}{\end{eqnarray}}

\newcommand{\pb}[1]{\hbox{\lower0.5ex\hbox{${}_{\leftarrow}$}}\kern-1.9ex{#1}}

\def\be{\begin{equation}}
\def\ee{\end{equation}}
\def\ba{\begin{eqnarray}}
\def\ea{\end{eqnarray}}

\def\h{\hat}

\def\f{\frac}

\def\rmd{\mathrm{d}}

\def\F{\mathcal{F}}
\def\A{\mathcal{A}}
\def\H{\mathcal{H}}
\def\W{\mathcal{W}}
\def\g{\mathfrak{g}}
\def\E{\mathcal{E}}

\def\hphi{\hat{\phi}}
\def\hPhi{\hat{\Phi}}
\def\stress{\langle \hat{T}_{ab}(x) \rangle_{\rm ren}}
\def\phisq{\langle \hat{\phi}^2(x)\rangle_{\rm ren}}

\def\Mo{\mathring{M}}

\def\go{\mathring{{g}}}
\def\phio{{\phi}^\circ}

\def\hphio{{\hat{\phi}^\circ}}
\def\biphi{\langle \hphi(x)\,\hphi(\xp)\rangle}
\def\biphio{\langle \hphio(x)\,\phio(\xp)\rangle}

\def\Boxo{\mathring{\Box}}
\def\Fo{\mathring{F}}

\def\hPhi{\h\Phi}

\def\pso{\mathring{\Gamma}_{\rm cov}}

\def\Vo{\mathring{V}}

\def\ps{\Gamma_{{\rm Cov}}}

\def\vx{\vec{x}}
\def\vk{\vec{k}}

\def\K{\sqrt{K}}

\def\xp{x^{\,\prime}}

\begin{document}
\title{\bf Probing cosmological singularities with quantum fields:\\ Open and closed FLRW universes}

\author{Abhay Ashtekar}\email{ashtekar.gravity@gmail.com}
\affiliation{Institute for Gravitation and the Cosmos \& Physics Department,The Pennsylvania State University, University Park, PA 16802 U.S.A}
\author{Adri\'an del R\'io}\email{axd570@psu.edu}
\affiliation{Institute for Gravitation and the Cosmos \& Physics Department,The Pennsylvania State University, University Park, PA 16802 U.S.A}

\begin{abstract}

It was recently pointed out that linear quantum fields $\hphi(x)$ can be meaningfully propagated across the big bang (and the big crunch) singularities of spatially flat Friedmann, Lema\^itre, Robertson, Walker (FLRW)universes \cite{ADLS2021}. Recall that $\hphi(x)$, as well as renormalized observables $\phisq$ and $\stress$, are distribution-valued already in Minkowskian quantum field theories.  It was shown that they can be extended as well-defined distributions even when these space-times are enlarged to include the big-bang (or the big crunch). We generalize these results to spatially closed and open FLRW models, showing that this `tameness' of cosmological singularities is not an artifact of the technical simplifications due to spatial flatness. Our analysis also provides explicit expressions of $\biphi,\, \phisq$ and $\stress$ in closed and open universes for minimally coupled massless scalar fields  and discuss the ambiguities in the definition of $\stress$ at the big-bang. While the technical expressions are more complicated than in the spatially flat case, there is also an unexpected conceptual simplification: the infrared divergence \cite{fp} is now absent because, in effect, the spatial curvature provides a natural cutoff. Finally, we further clarify the sense in which quantum field theory can continue to be well defined even though the extended space-time is not globally hyperbolic because of the singularity, and suggest directions for further work.

 \end{abstract}

\date{\today}
\maketitle

\section{Introduction}
\label{s1}

Singularity theorems of Penrose, Hawking, Geroch and others are often interpreted to mean that the big bang (and the big crunch) represent the absolute beginning (and end) of the universe: since these singularities are space-like, space-time is taken to terminate there and it is assumed that physics cannot be extended beyond. However, the theorems only establish geodesic incompleteness. While trajectories of classical test particles do come to an abrupt end, they are not the appropriate tools to probe the space-time structure once the curvature enters the Planck regime. 
It has been long argued that singularities of general relativity may be quite tame when probed with more realistic tools. For example, quantum particles were used as probes in certain static, dilatonic black holes in Ref.~\cite{HM1995}, and classical test fields were used as probes in superextremal Reissner-Nordstr\"om solutions in Ref.~\cite{STZ2004}. In these cases, the background space-times are non-dynamical and the singularities are eternal. The conclusion was that the evolution of these probes remains well-defined in a precise mathematical sense in spite of the singularity. In this paper we will focus on cosmological space-times which are \emph{dynamical} and in which the singularity is physically more interesting. Because of the dynamical nature of space-time, there is `particle creation' and quantum test particles are no longer suitable as probes. One has to use quantum fields which in turn require appropriate mathematical tools to handle the presence of an infinite number of degrees of freedom. The simplest setting is provided by the spatially flat (or $K=0$) Friedmann, Lema\^itre, Robertson, Walker (FLRW) models. Linear, test quantum fields were recently analyzed on this background \cite{ADLS2021}. A key point in this analysis was to note that, already  in Minkowski space-time, a quantum field $\hphi(x)$ is an operator-valued \emph{distribution} (OVD) on the Fock space, rather than an operator. Indeed, it is because of this distributional character that one needs a regularization and renormalization procedure to define their products such as $\phisq$. One cannot expect $\hphi(x)$ to do better at the big-bang singularity! Now, in FLRW models we can extend the space-time across the big-bang (and/or big-crunch) singularity in an obvious way so that the space-time metric $g_{ab}$ is only continuous (and degenerate) at the singularity. The questions then are:\\
(i) Can  $\hphi(x)$ be extended as a well-defined OVD in the larger space-time?\,\\ 
(ii) Can the associated $\biphi$ be extended as a well-defined  bi-distribution? and, \\
(iii) Can observables $\phisq$ and $\stress$ be extended as well-defined distributions?\\
In the $K=0$ case, if we consider scale factors of the form $a(\eta) = a_\beta \eta^\beta$ (where $\eta$ is the conformal time), we have $\beta =1$ for radiation-filled universes and $\beta=2$ for dust-filled universes. In the main text of \cite{ADLS2021} the focus was on these two cases and the situation for $\beta > 2$ was summarized in an Appendix. Somewhat surprisingly, the answers to all three questions turned out to be in the affirmative. 
\footnote{Another example of direct physical interest is provided by the Schwarzschild space-time in which the singularity is also space-like and in its vicinity space-time is again dynamical. This has been analyzed using a formal Schr\"odinger representation in \cite{HS2015}, and at a mathematical level that pays due attention to the presence of an infinite number of degrees of freedom in \cite{AS2022}. (Bianchi I model has also been investigated using the formal Schr\"odinger representation in \cite{HSU2019}). The conclusion is again that, when probed with linear, test quantum fields, the singularity is tame.}

Findings in \cite{ADLS2021} were facilitated by the technical simplifications associated with the absence of spatial curvature in the $K=0$ case. The question is whether the final results depend critically on those simplifications. If so, regularity of quantum fields across the big-bang would be accidental. To gain insight into this issue, in this paper we will consider open and closed FLRW universes which are technically more complicated due to the presence of spatial curvature. We will focus on radiation and dust filled universes and show that the main results of \cite{ADLS2021} remain unaltered, although the expressions of $\biphi$, $\phisq$ and $\stress$ are now more complicated. (More general equations of state for matter will not be discussed because the analysis becomes technically even more complicated and findings of \cite{ADLS2021} suggest that the main conclusions will not change qualitatively.) Somewhat surprisingly --at least at first-- the presence of spatial curvature leads to a \emph{conceptual simplification}. In the $K=0$ case, there is an infrared (IR) divergence for dust-filled universes that requires the introduction of an IR regulator \cite{fp}. We will find that this is no longer necessary in closed and open universes because spatial curvature introduces an effective IR cut-off. We will also take this opportunity to add some clarifications about the effect of the singularity. Specifically, because the extended space-time that includes the singularity is no longer globally hyperbolic, some of the familiar results fail to go through. Nonetheless, there is sufficient structure to enable one to introduce the OVD $\hphi(x)$ and investigate its properties on the extended space-time that includes the singularity.

The material is organized as follows. In Sec.~\ref{s2} we briefly review the theory of linear quantum fields in FLRW space-times. This discussion will serve to fix the notation and also enable us to separate the structure that fails to go through when the space-time is extended to include the big-bang, from that which does go through. Relative to the $K=0$ case, the main technical complication is that, whereas the eigenfunctions of the spatial Laplacian are simply the plane waves $e^{i\vk\cdot\vx}$ in the $K=0$ case, they are much more involved in presence of spatial curvature. In Sec.~\ref{s3} we extend the space-time across the big-bang/big-crunch of the radiation-filled universes, and show that sufficient structure continues to be available to establish that the OVD $\hphi(x)$ is well-defined on the extension. As in the $K=0$ case, the time-dependence of mode functions that define the explicit form of $\hphi(x)$ in the standard Fock representation is quite simple because the space-time scalar curvature $R$ vanishes also for closed and open  universes. We will find that the bi-distribution $\biphi$ continues to have the standard Hadamard structure away from the big-bang/big-crunch surface, and if the points $x,\,x^\prime$ have a `purely time-like' separation, also across this surface. Using this bi-distribution we then calculate the observables $\phisq$ and $\stress$ using Hadamard renormalization. In Sec.~\ref{s4} we turn to dust-dominated universes. Now the space-time scalar curvature $R$ no longer vanishes, whence the time dependence of mode functions is more complicated. Still, one can carry out all the required calculations and again obtain the explicit expressions of $\biphi, \phisq$ and $\stress$ as in Sec.~\ref{s3}. It is manifest that the IR cut-off that is necessary in the $K=0$, dust-filled FLRW universe is no longer necessary in closed and open universes. For both radiation and dust filled universes, the final expressions of $\phisq$ and $\stress$ make it manifest that these observables are well defined distributions on the full, extended space-time in spite of the singularity. This can be quite surprising to researchers who focus on cosmological perturbations and relation between theory and observations. However, we recently learned that this fact was expected by some in the algebraic quantum field theory and microlocal analysis community, based on results of \cite{steinmann,bf}. In sections \ref{s3} and \ref{s4} we also provide explicit expressions based on the Hadamard renormalization scheme that would be of direct interest to semi-classical gravity. Our expressions of $\stress$ satisfy the standard properties \cite{rmw-book}, including the non-trivial conservation requirement.  But still the answer is as usual scheme dependent and there is  a further 4-parameter ambiguity in the definitions of these distributions at the big-bang/big-crunch. Since our primary goal is to show existence rather than uniqueness, we choose a commonly used prescription motivated by certain mathematical properties to fix them. Whether they can be fixed using physical criteria is an interesting open issue.

The presence of the singularity in the extended space-time does give rise to certain subtleties. In particular, since the extended space-time is not globally hyperbolic, the standard notion of retarded and advanced Green's functions are not available, and the standard notion of Hadamard regularity cannot be applied if the points lie on the big-bang/big-crunch surface. This is not surprising. Rather, what is surprising is that these issues do not pose unsurmountable obstacles. At appropriate points in sections~\ref{s3} and \ref{s4}, we explain how apparent difficulties are circumvented. In Sec.~\ref{s5} we summarize the main results, provide a general perspective on the quantum tameness of space-like singularities of classical general relativity, and suggest directions for further work.

One of the primary goal of this paper is to try to bridge the gap between the community that investigates quantum field theory using mathematically rigorous techniques, generally emphasizing the algebraic approach, and the community that is primarily interested in applications of the theory to the early universe, emphasizing explicit expressions of $\hphi(x)$ and $\biphi$ using mode expansions and of $\phisq$ and $\stress$ using specific renormalization schemes, in the hope that combined ideas from both communities will shed further light on the nature of the big bang when probed using quantum tools. To make the material accessible to both communities, we have tried to strike a balance between mathematical rigor and physical and conceptual issues.

Our conventions are as follows. We work in the geometrized units $G=c=1$. The metric signature has signature $(-,+,+,+)$, the Riemann tensor is defined by\, $2 \nabla_{[a}\,\nabla_{b]}v_c=:R_{abc}{}^{d}v_d$\, for any 1-form $v_d$; the Ricci tensor is  defined as $R_{ab}:=R_{acb}{}^c$; and the scalar curvature as $R:=g^{ab}R_{ab}$. In section~\ref{s2} we work with the standard FLRW models that exclude the big-bang and big-crunch singularities, and all classical fields are assumed to be smooth. In sections~\ref{s3} and ~\ref{s4}, the space-time manifold is enlarged to include the big-bang/big-crunch by extending the range of conformal time $\eta$ in the obvious manner. Then, the inverse scale factor, curvature tensor and classical solutions to the Klein-Gordon equation are treated as distributions which are smooth fields away from the singularity. 

\goodbreak
%%%%%%%%%%%%%%%%
\section{Linear quantum fields in FLRW space-times}
\label{s2}
%%%%%%%%%%%%%%%%

In this section we restrict ourselves to the FLRW space-times that are smooth, i.e., exclude the big-bang and big-crunch singularities. Discussion is divided into three parts. In the first, we recall structures underlying quantum field theory in curved space-times \cite{am,rmw-book,bfv,BDH,hw,cfkr,gns1,gns2}
that are needed for our analysis, and in the second, the basics of mode decomposition of scalar fields in spatially closed and open FLRW models \cite{eigenvalues}. This summary will help make the paper self-contained. To keep the summary reasonably short, and to reach the audience outside the `communications in mathematical physics community', we will have to gloss over issues that are not essential for our main results, such as those related to topology on infinite dimensional spaces. Therefore, we will not use notions from $C^\star$-algebras, nor discuss functional analytic issues such as the distinction between weakly non-degenerate and strongly non-degenerate symplectic structures. In the third part we use the results from the first two parts to arrive at a strategy that will facilitate the extension of quantum fields across the big-bang/big-crunch. 

\subsection{Operator algebras and their quasi-free representations}
\label{s2.1}

Consider a scalar field $\phi(x)$ satisfying the Klein-Gordon equation $(\Box - m^2 - \xi R)\,\phi(x) =0$ on a globally hyperbolic space-time $(M, g_{ab})$ where $m, \xi$ are constants and $R$ denotes the scalar curvature of the space-time metric $g_{ab}$. Since  the Klein-Gordon operator is normally hyperbolic, we have well-defined retarded, advanced and the causal/commutator Green's functions, $G_{\rm Ret}(x,\,x^\prime),\, G_{\rm Ad}(x,\,x^\prime),\, \Delta (x,\,x^\prime)$, respectively, with $\Delta (x,\,x^\prime) = (G_{\rm Ad}\, -\, G_{\rm Ret})(x,\,x^\prime)$ \cite{lichne}. Then, given any $C^\infty$ function $f(x)$ with compact support on $M$, \, i.e., an element of $C^\infty_0(M)$,  
\be \label{sol1} F(x) := \int_M \rmd^4 V^\prime  \Delta(x,\, \xp) f(\xp) \ee
is a smooth solution to the Klein-Gordon equation, whose initial data is of compact support on any Cauchy slice. 
(Here $\rmd^4 V^\prime$ is the volume element of the metric $g_{ab}$ on $M$.) In fact, any smooth solution $F$ of the Klein-Gordon equation that has spatially compact support  on every Cauchy slice is of this form for some $f\in C^\infty_0(M)$. Note, however, that the map $f(x) \to F(x)$ has a large kernel: If $f(x) = (\Box - m^2 - \xi R) h(x)$ for \emph{any} $h(x) \in C^\infty_0(M)$, then the corresponding solution is identically zero.  Now, the space of these solutions $F(x)$ is naturally equipped with a symplectic structure $\Omega$ that defines Poisson brackets in the classical theory and leads to the canonical commutation relations in the quantum theory:
\be \label{Omega} \Omega(F_1(x),\, F_2(x)) := \int_{\Sigma}\!\! \rmd^3 V \big[F_1(x) (n^a\nabla_a \,F_2(x) -  (n^a \nabla_a F_1(x)) F_2(x))\big] \, \ee
where $\Sigma$ is any Cauchy surface in $M$ and $n^a$ the unit normal to $\Sigma$. Satisfaction of the Klein-Gordon equation guarantees that the integral on the right side is independent of the choice of $\Sigma$. This fact will play an important role in our extension of the theory across the big-bang/big-crunch. We can re-express the symplectic structure using any of the test functions $f_1(x)$ and $f_2(x)$ that give rise to $F_1(x)$ and $F_2(x)$ via (\ref{sol1}):
\be \label{Omega2} \Omega(F_1(x),\, F_2(x)) \,=\, \int_M\! {\rmd}^4 V 
    \int_M {\rmd}^4 V^\prime \,\Delta (x,\,\xp)\, f_1(x)\, f_2(\xp)\, . \ee

Now, in the algebraic approach to quantum field theory in curved space-times, one generally begins by introducing an (abstractly defined) OVD $\hphi(x)$ that satisfies: \\
(i) $\star$-relations: $\hphi(f) = \hphi^\star(f)$, where $\hphi(f)= \int_M \rmd^4 V \,\hphi(x) f(x)$;\\
(ii) The Klein Gordon equation in a distributional sense: 
\be \label{fe}  \hphi ((\Box-m^2- \xi\,R)f) := \int_M \!\rmd^4 V\, \hphi(x)\, (\Box - m^2 -\xi\,R) f(x) =0\, ; \ee 
(iii) The commutation relations:
\be \label{comm2}
    [\hphi(f_1),\, \hphi(f_2) ] = \,{i\hbar}\int_M\! {\rmd}^4 V f_1(x)
    \int_M {\rmd}^4 V^\prime \,\Delta (x,\,\xp)\,  f_2(\xp)\,\,\h{I} , \ee
for any $f, f_1, f_2$ in $C^\infty_0(M)$. One then constructs the free $\star$-algebra $\mathcal{A}_{(f)}$ generated by the operators $\hphi(f)$ subject to these three relations. Note that map $f \to \hphi(f)$ has, again, a large kernel because of (\ref{fe}). Therefore, it is often convenient to pass from $f(x) \in C^\infty_0(M)$ to $F(x)$ and associate field operators $\hPhi(F)$ with suitably regular classical solutions $F(x)$, defined formally as follows:
\be \hPhi(F) = \Omega(\hphi(x),\, F(x)) \equiv \int_{\Sigma}\!\! \rmd^3 V \big[\hphi(x) (n^a\nabla_a \,F(x) -  (n^a \nabla_a \hphi(x)) F(x))\big] \, . \ee
We can now construct a $\star$-algebra $\mathcal{A}_{(F)}$ generated by $\hPhi(F)$. Since $F(x)$ is a smooth  solution to the KG equation, we no longer need the condition (ii) above: $\hPhi(F)$ are subject only to\\    
($i^\prime$) $\star$-relations: $\hPhi(F) = \hPhi^\star(F)$; and,\\
($iii^\prime$) Commutation relations \be \label{comm}  [\hPhi(F_1),\, \hPhi(F_2) ] = \,{i\hbar}\, \Omega (F_1, F_2)\, \h{I}\, .\ee
Then, in place of $\mathcal{A}_{(f)}$, we can use the $\star$-algebra $\mathcal{A}_{(F)}$ generated by the operators $\hPhi(F)$ (defined, again, abstractly), subject to ($i^\prime$) and ($iii^\prime$) \cite{am}. The map $\hphi(f) \to \hPhi(F)$ --where $f$ determines $F$ via (\ref{sol1})-- is an isomorphism from the $\star$-algebra $\mathcal{A}_{(f)}$ to the $\star$-algebra $\mathcal{A}_{(F)}$. We will find that $\mathcal{A}_{(F)}$ is particularly useful in extending the quantum field theory beyond the big-bang because, e.g., this algebra is not tied to the use of Green's functions, which refer to global hyperbolicity.

Recall that the Gel'fand, Naimark, Segal (GNS) construction \cite{gns1,gns2}  provides a direct and elegant avenue to obtain representations of $\star$-algebras by operators on a Hilbert space. Given a normalized, positive linear function (PLF) on a $\star$-algebra --in physical terms, an expectation value function $\E$-- the construction provides an explicit, step by step procedure to build a Hilbert space and represent elements of the algebra by concrete operators on it such that all algebraic relations are preserved. To specify PLFs $\E$, it is much more convenient to work with the Weyl operators $\h{W}(F) := \exp \f{i}{\hbar} \hPhi(F)$ because the vector space of their linear combinations is closed under the product:
\be \label{prod} \h{W}(F_1) \h{W}(F_2) = e^{-\f{i}{2\hbar} \Omega(F_1, F_2)}\, \h{W}(F_1+F_2)\, .\ee
The \emph{Weyl algebra} $\W$, generated by $\h{W}(F)$ is thus spanned just by linear combinations of $\h{W}(F)$.%
\footnote{$\W$ can be endowed with the structure of a $C^\star$ algebra in a natural manner. It is more appropriate  to use it in the GNS construction, in particular in the discussion of self-adjointness of field operators $\hPhi(F)$.} 
Because of this property (\ref{prod}), the task of specifying a PLF $\E$ on the Weyl algebra $\W$ simplifies considerably, since one only has to specify $\E(W(F))$. (By contrast, on $\mathcal{A}_{(F)}$, for example, one has to specify $\E(\hPhi(F_1) \ldots \hPhi(F_n))$ for all $n$.) 

Of particular interest are the quasi-free (or, Gaussian) states that lead to Fock representations of these $\star$-algebras, in which $\hPhi(F)$ (and $\hphi(f)$) are represented as (unbounded) sums of concretely defined creation and annihilation operators, and $\h{W}(F)$ by the corresponding unitary (and therefore bounded) operators. Let us denote by $\ps$ the space of suitably regular solutions $F(x)$ of the Klein-Gordon equation. Then, choice of a quasi-free state on $\W$ is in 1-1 correspondence with the choice of a positive definite metric $\g$ on $\ps$ which is compatible with the symplectic structure $\Omega$ on $\ps$. Given such a $\g$, 
\be \label{plf}\E(\h{W}(F)) := e^{-{\f{1}{4\hbar}}\, \g(F,F)}\,  \ee
naturally extends to a positive-linear function on all of $\W$. In the resulting GNS representation, $\g$ is (essentially) the `covariance matrix' of the vacuum state. The compatibility condition between $\g$ and $\Omega$ is that $\g$ is of the form $\g(F_1, F_2) := \Omega(F_1, JF_2)$, where $J$ is a complex structure, i.e., a linear operator on $\ps$ satisfying $J^2 = -1$. The triplet $(\Omega, g, J)$ endows $\ps$ with a K\"ahler structure for which the Hermitian inner product is given by
\be \label{ip1} \langle F_1,\,  F_2\rangle = \f{1}{2\hbar}\, \big(\g(F_1,\, F_2) + i\, \Omega (F_1,\, F_2) \big)\,  \ee 
(so that the PLF has the familiar form $\E(W(F)) := e^{-\f{1}{2} \langle F\,|\, F\rangle}$). The Cauchy completion of $\ps$ with respect to this inner product is the 1-particle Hilbert space $\H$ in the quasi-free representation defined by (\ref{plf}). The full Hilbert space determined by the GNS construction is the symmetric Fock space $\F$ based on $\H$. The representation map $\pi$ --determined by the choice of the metric $\g$ or the complex structure $J$-- sends the abstractly defined field operator $\hPhi(F)$ to the concrete operator on $\F$,
\be
\pi(\hat\Phi(F)) = \hbar[ \h{A}(F)+\h{A}^\dag(F) ]\, ,  \label{repquantumfield}
\ee
where the concrete creation and annihilation operators $A^\dag(F)$ and $A(F)$ satisfy the usual commutation relations:
\be
[\h{A}(F_1),\, \h{A}^\dag(F_2)]\,=\,  \langle F_1,F_2\rangle \, \h{I}\, .
\ee
For simplicity, from now on we will drop the explicit representation map $\pi$; the context will make it clear whether we are referring to abstract operators of their concrete representations on $\F$.

We will see in sections \ref{s3} and \ref{s4} that one can naturally extend the familiar phase spaces $\ps$  associated with the post big-bang FLRW space-times  --together with the K\"ahler structure $(J,\, \g,\, \langle\, .\, |\, .\,\rangle)$ they carry-- across the big-bang. This will provide an extension of standard quantum field theories to larger space-times in spite of the singularity.

\subsection{Klein-Gordon fields in FLRW cosmologies}
\label{s2.2}

We will now turn to FLRW cosmologies and summarize certain structures underlying classical Klein Gordon fields on these space-times that will be used in Sec.~\ref{s3} and \ref{s4} to construct $\star$-algebras and quasi-free representations. 

For models of interest to this paper, the space-time metric $g_{ab}$ is given by %
\be  \label{metric} {g}_{ab}\rmd x^{a} \rmd x^{b} =:  a^{2}(\eta)\, \go_{ab}\rmd x^{a} \rmd x^{b} \equiv
a^{2}(\eta) \big(-\rmd \eta^{2} + h_{ij} \rmd x^i\, \rmd x^j\big), \ee
where $\eta$ is the conformal time and the spatial metrics $h_{ij}$ are of constant curvature $K$.  Since $K$ is the scalar curvature of the 3-manifold, it has dimensions of $1/({\rm length})^2$. Following the usual conventions, we will set 
\be \K = |\sqrt{K}|\,\,\, {\hbox{\rm if $K >0$}} \qquad {\rm and}\qquad \sqrt{K} = i |\sqrt{K}|\,\,\,
{\hbox{\rm if $K <0$}} \ee
and, of course, $\K=0$ if $K=0$. Then, $h_{ij}$ has the explicit form
\be h_{ij} \rmd x^i\, \rmd x^j = \rmd \chi^2 \, +\, \Sigma^2_{(K)}(\chi)\,\, (\rmd \theta^2 + \sin^2\theta\rmd \varphi^2) \quad {\rm where} \quad \Sigma_{(K)}(\chi) = \f{\sin \K\chi}{\K} \ee
The domain of the radial coordinate $\chi$  is $(0,\infty)$  for both $K=0$, $K <0$, and $\chi\in(0,\pi)$ for  $K>0$, while $(\theta,\phi)$ are the usual coordinates on the 2-sphere. The scalar curvature $R$ of $g_{ab}$ is given by $R(\eta) = \,\f{6}{a^2(\eta)} \, \big(\f{a^{\prime\prime}}{a} + K)$  where the prime denotes derivative w.r.t. $\eta$. Note that that the 4-metric $\go_{ab}$ defined in (\ref{metric}) is ultra-static --it admits $\partial/\partial \eta$ as a hypersurface-orthogonal, time-like Killing vector with constant norm-- but it is not flat unless $K=0$.

Because of the form of the metric $g_{ab}$, without loss of generality we can separate the spatial and time dependence of solutions to the Klein-Gordon equation $(\Box\, - m^2-\, \xi R)\phi =0$ and  consider solutions of the form 
\be \label{basis}
\phi_{\vec k}(\eta,\vec x) = \frac{\sqrt{\hbar}}{a(\eta)}\,\,\, \psi_k(\eta)\,\, \mathcal{Y}^{(K)}_{\vec k}(\vec x)
\ee
where $\vx \equiv (\chi,\theta,\phi)$ and $\vec k$ labels the eigenfunctions of the spatial Laplacian $\Delta^{(K)}$ defined by $h_{ij}$. As in the spatially flat case, the solution diverges when the scale factor $a(\eta)$ vanishes. With this ansatz, the Klein-Gordon equation separates into two parts, the first dictating the spatial dependence and the second governing the time evolution:
\bea
\bigtriangleup^{(K)}\, \mathcal Y^{(K)}_{\vec k}(\vec x) +(k^2- K)\,\mathcal Y^{(K)}_{\vec k}(\vec x) &  = & 0\, ,\quad {\rm and}   \label{spatialeq}\\
\psi_k^{\prime\prime}+\Big(k^2+m^2 a^2+\big(\xi-\frac{1}{6}\big)a^2 R\Big)\,\psi_k & = & 0\, , \label{timeeq}
\eea
where the prime denotes derivative w.r.t. $\eta$ in (\ref{timeeq}), and $k$ is the separation constant. Eq. (\ref{spatialeq}) is just the eigenvalue equation for the spatial Laplacian $\Delta^{(K)}$. For $K=0$, the solutions are just plane waves $e^{i\vk\cdot\vx}$. But for non-zero $K$, the solutions are more complicated. Nonetheless, one can analyze their structure in detail because the spatial metric $h_{ij}$ is of constant curvature  both for $K >0$ and $K <0$ (see Appendix~\ref{a2}).\, However, the time evolution equation (\ref{timeeq}) does not share this `universality': the form $a(\eta)$ of the scale factor varies from one model to another depending on matter content, and the equation also involves the mass $m$ and the conformal coupling constant $\xi$. Therefore, in the detailed analysis of sections \ref{s3} and \ref{s4} we will set $m=0$ and $\xi=0$ --thus focusing on the massless, minimally-coupled Klein-Gordon equation-- and restrict ourselves to scale factors $a(\eta)$ corresponding to radiation and dust-filled universes. 
In the remaining part of this sub-section, we will focus on the spatial equation (\ref{spatialeq}).

Main properties of eigenvalues and eigenfunctions of $\Delta^{(K)}$ can be summarized as follows. First, because the 3-metric $h_{ij}$ is spherically symmetric, we can separate the eigenfunctions 
$\mathcal Y^{(K)}_{\vec k}(\vec x)$ into a radial and a spherical part:
\be 
\mathcal Y^{(K)}_{\vec k}(\vec x) =  \Pi^{(K)}_{k\ell}(\chi)\,\,Y_{\ell m}(\theta,\phi)\, , \label{decom}
\ee
where $Y_{\ell m}(\theta,\phi)$ are the usual spherical harmonics, with $\ell=0,1,2, \dots$, and $m=-\ell,-\ell+1,\dots, \ell$. Thus, the label $\vk$ of the eigenfunctions of the Laplacian $\Delta^{(K)}$ stands for the triplet $k, \ell, m$.
The structure of the radial modes $\Pi^{(K)}_{k\ell}(\chi)$ is rather complicated and is discussed in detail in the Appendix~\ref{a2}. Here we will simply note the properties that are important to our analysis. The general solution to the radial equation is given by
\ba
 \Pi^{(K)}_{k\ell}(\chi)= A_{k \ell}\, \sin^{\ell}({\K}\chi) \left[\frac{\rmd}{\rmd\cos(\K\chi)} \right]^{\ell+1}\!\!\!\!\cos(k\chi)+B_{k \ell}\sin^{\ell}(\K\chi) \left[\frac{\rmd}{\rmd\cos(\K\chi)} \right]^{\ell+1}\!\!\!\!\sin(k\chi) \label{radialmodes}\nonumber\\
\ea
where $ A_{k \ell}$, $B_{k \ell}$ are arbitrary coefficients. In the $K \to 0$ limit, the two linearly independent solutions reduce to the usual spherical Bessel functions $j_{\ell}(k\chi)$, and the Neumann functions $y_{\ell}(k\chi)$, respectively (see, for instance, Eqs. (10.1.25) and (10.1.26) of \cite{ASbook}). As one would expect from this limit, regularity conditions at the origin $\chi=0$ require that we set $B_{k\ell} =0$ for all $k, \ell$ and $K$. The resulting functions are even in $k$, so without loss of generality we will restrict ourselves to $k>0$ (note that solutions vanish for $k=0$). Furthermore,  for $K >0$, a detailed examination shows that regularity on the entire spatial manifold $\mathbb{S}^3$, including points $\chi=\pi$, imposes the following additional conditions:\\
(i) $k$ is quantized:\,\, $k = \K \,n \quad {\rm where}\,\,  n$ \, is a positive integer, and \\
(ii) For any given $k \ge \K$, the quantum number $\ell$ is bounded above: it can only take values: $\ell = 0, 1, ..., \big((k/\K) -1\big)$. \\
For the case when the spatial manifold is a hyperboloid  $\mathbb{H}$, we have $ K <0$ and there is no restriction either on $k$ --which takes values in $(0, \infty)$-- or, on permissible values of $\ell$.\vskip0.2cm

Finally, let us choose the coefficients $A_{kl}$ so as to normalize the eigenfunctions appropriately. We will set 
\be \label{Pikl}
\Pi^{(K)}_{k\ell}(\chi)=\frac{\K}{\sqrt{\frac{\pi}{2}  \Pi_{n=0}^{\ell} [k^2/K - n^2] }  } \sin^{\ell}(\K\chi) \left[\frac{\rmd}{\rmd\cos(\K\chi)} \right]^{\ell+1}\cos(k\chi)
\ee
Note that this expression is well-defined for $k=0$ for any $\ell$. This choice ensures (1) the orthonormality condition 
\be \label{ortho1}
\int_{\Sigma} \rmd^3\Vo\,\,  \mathcal Y^{(K)}_{\vec k}(\vec x) \, \bar{\mathcal Y}^{(K)}_{\vec k'}(\vec x)\,=\,\delta(k-k')\,\delta_{\ell,\ell'}\,\delta_{m,m'} 
\ee
for $K\le0$, where $\rmd^3\Vo$ is the volume element defined by the spatial metric $h_{ij}$;  (2) the orthonormality condition 
\be 
\int_{\Sigma} \rmd^3\Vo\,\,  \mathcal Y^{(K)}_{\vec k}(\vec x) \, \bar{\mathcal Y}^{(K)}_{\vec k'}(\vec x)\,=\,
\f{1}{\sqrt{K}}\,\delta_{k,k^\prime}\,\delta_{\ell,\ell'}\,\delta_{m,m'} \ee
for $K>0$; and, (3) the completeness relation
\be \label{completeness}
\sum_{\ell m} \mathcal Y^{(K)}_{\vec k}(\vec x)\,\, \mathcal {\bar Y}^{(K)}_{\vec k}(\vec x')\, =\, \sum_{\ell m}  \Pi^{(K)}_{k\ell}(\chi)\,  \bar\Pi^{(K)}_{k\ell}(\chi')\,\, |Y_{\ell m}(\theta,\phi)|^2 = \frac{k}{2\pi^2} \frac{\sin[k(\chi-\chi')]}{\Sigma_K(\chi-\chi')}\, . 
\ee

\subsection{Quantization Strategy}
\label{s2.3}

Recall from Sec.~\ref{s2.1} that one can construct the algebra $\A_{(F)}$ of operators and its quasi-free representations if one has the following ingredients: (i) A space $\ps$ of suitably regular solutions $F(x)$ to the field equation; and, (ii) a complex structure $J$ thereon that is compatible with the symplectic structure $\Omega$ so that (\ref{ip1}) is an Hermitian inner product on $\ps$. 
This formulation was shown in \cite{ADLS2021} to be well-suited for extending the theory across the big-bang/big-crunch in the $K=0$ case. In this section we will explain why the same strategy can be successfully used in closed and open FLRW models.

Let us begin with FLRW space-times that are smooth and exclude singularities. Then, the first step in our construction would be to pick a complete set of candidate `positive frequency solutions'. A standard practice accomplishes this by specifying a basis $e_k(\eta)$ in the space of solutions to the time evolution equation (\ref{timeeq}) with following two properties: \\
(i) For each $k$, the $e_k(\eta)$, together with their complex conjugates, span the solution space; and, \\
(ii)They are normalized such that 
\be \label{normalization} e_k(\eta)\, \partial_\eta \bar{e}_k(\eta) -  \bar{e}_k(\eta)\, \partial_\eta {e}_k(\eta) = i\, . \ee
Then the required $\ps$ can be constructed by taking suitable linear combinations of $(e_k(\eta)/a(\eta)) \, \mathcal{Y}_{\vk}^{(K)} (\vx)$ (see Eq.~(\ref{basis})). Coefficients in these linear combinations --denoted by $z(\vk)$ below-- have to be specified with due care to ensure that the operators $\hPhi(F)$ are well-defined on the resulting Fock representation, and continue to be well-defined even when the FLRW metric is extended across the big-bang. We will choose them as follows. Consider $C^\infty$ functions $z(\vx)$ of compact support on the spatial manifolds of constant curvature (namely, the $\eta={\rm const}$ surfaces) and denote by $z(\vk)$ their expansion coefficients in the normalized basis $\mathcal{Y}_{\vk}^{(K)} (\vx)$.  In the $K=0$ case, these $z(\vk)$ become the Fourier transforms of $C_0^\infty$ functions $z(\vx)$ on the Euclidean space. As in that case, the functions $z(\vk)$ fall-off in $k$ faster than any polynomial as $k \to \infty$ ensuring convergence of various $k$ integrals (or sums). 

Let us now consider the $K<0$ case. Then, our $\ps$ will consist of solutions $F$ to the Klein-Gordon equation of the form
\ba \label{F} F(\eta, \vx) &=&  \f{1}{a(\eta)}\, \int_0^\infty \rmd k \sum_{\ell=0}^\infty\, \sum_{m=-\ell}^{\ell} \Big[z(\vk)\, e_k(\eta)\, \mathcal{Y}^{(K)}_{\vk} (\vx)\, +\, \bar{z}(\vk)\, \bar{e}_k(\eta)\, \bar{\mathcal{Y}}^{(K)}_{\vk} (\vx)\, \Big] \\
&=:& F^+(\eta, \vx) + F^-(\eta, \vx)
\ea
(see Eq.~(\ref{basis})).\, $F^+(\eta,\vx)$\, --the first term on the right side of (\ref{F})-- \,is to be thought of as the `positive frequency part' of $F(\eta,\vx)$, and its complex conjugate, the `negative frequency part'. The normalization conditions on $e_k(\eta)$ and $\mathcal{Y}^{(K)}_{\vk} (\vx)$ then ensure that the linear operator $J$ defined on $\ps$ by 
\be (JF)(\eta, \vx) := i\,F^+(\eta, \vx)\, -\, i\, F^-(\eta, \vx) \label{J}\ee
is a complex structure on $\ps$ that is compatible with the symplectic structure (\ref{Omega}) thereon. The resulting positive definite K\"ahler metric $\g$ they define is given simply by
\be \g(F,\, F) :=\, 2\int_0^\infty \rmd k \sum_{\ell=0}^\infty\, \sum_{m=-\ell}^{\ell} |z(\vk)|^2 \, . \label{norm}
\ee
As noted in Sec.~\ref{s2.1}, $\g$ naturally provides a PLF on the Weyl algebra $\W$ and carries the interpretation of the `covariance matrix' of the vacuum in the resulting Fock representation. On this Fock space,  the OVD $\hphi(x)$ admits an expansion that mimics (\ref{F}): 
\be \label{hphi} \hphi(\eta, \vx) =  \f{1}{a(\eta)}\, \int_0^\infty \rmd k \sum_{\ell=0}^\infty\, \sum_{m=-\ell}^{\ell} \Big[e_k(\eta)\, \mathcal{Y}^{(K)}_{\vk} (\vx)\, \h{A}(\vk)\, +\, \bar{e}_k(\eta)\, \bar{\mathcal{Y}}^{(K)}_{\vk} (\vx)\, \h{A}^\dag(\vk) \Big]\ee
where the creation and annihilation operators satisfy  $[ \h{A}(\vk), \h{A}^\dag(\vk')]=\delta( k -  k')\delta_{\ell, \ell'}\delta_{m,m'}$.

Now a key point is that the time dependent part $\f{e_k(\eta)}{a(\eta)}$ that features in the expansion (\ref{F}) typically diverges at the big bang, $\eta=0$, first because the scale factor vanishes there, and second because in dust-filled FLRW universes $e_k(\eta)$ also diverges there. Therefore, the modes  --and hence $F(\eta, \vx)$-- are ill-defined as functions at $\eta=0$. However, as we will see in sections~\ref{s3} and \ref{s4}, the divergence is only polynomial, i.e., of the form $\eta^{-n}$. Now, we know from the standard distribution theory \cite{schwartz,gs,hormander,loja} that although $\eta^{-n}$ are singular as functions at $\eta=0$, they are well-defined as tempered distributions ${\underline{\eta}}^{-n}$ on the entire real line $\mathbb{R}$ that, furthermore, satisfy the familiar rules of calculus:%
\be\label{distribution} \f{\rmd}{\rmd \eta} {\underline{\eta}}^{-n} = - n\, {\underline{\eta}}^{-n-1} \quad {\hbox{\rm and,\quad if $n >1$\, then}}\quad  \eta\,\, {\underline{\eta}}^{-n} = {\underline{\eta}}^{-n+1}\, . \ee
(For a summary, see Appendix A of \cite{ADLS2021}.) Therefore if we extend the FLRW space-time across the big-bang by allowing $\eta$ to take values on the entire real line $(-\infty, \infty)$, then (${[e_k(\eta)}/{a(\eta)]}$\, and)\,  $F(\eta, \vx)$ are well-defined as distributions satisfying the Klein Gordon equation on the extended space-time.  

But, since the fields $F(x)$ diverge as functions at $\eta=0$, does the symplectic product $\Omega (F_1,\, F_2)$ not diverge there? It does not  simply because it is conserved. To obtain its explicit expression let us substitute the expansion (\ref{F}) in the expression (\ref{Omega}) of the symplectic structure and  evaluate it on a Cauchy surface $\Sigma$ given by $\eta=\eta_0 \not=0$. Then, using the orthonormality relations (\ref{ortho1}) and (\ref{normalization}) one finds
\be \Omega(F_1, \, F_2) \,=\, i \int_0^\infty \rmd k\, \sum_{\ell =0}^\infty\, \sum_{m=-\ell}^{\ell} \, \big [z_1(\vk) \bar{z}_2 (\vk) - z_2(\vk) \bar{z}_1(\vk)\big]\, . \ee
Since this expression is independent of the choice of $\eta_0$ we can extend it by continuity also to $\eta=0$ surface: the divergence of fields $F_1, F_2$ as we approach the big-bang is exactly compensated by the shrinking of the volume element \cite{ADLS2021}. As a consequence, the commutation relations (\ref{comm}) remain well-defined for any pair of solutions $F_1$, $F_2$ in $\ps$ even when the space-time is extended beyond the big-bang.
Therefore, we can construct the $\star$-algebra $\mathcal{A}_{(F)}$ generated by the operators $\hPhi(F)$ also on the extended space-time. We can also define a complex structure $J$ as in Eq.~(\ref{J}), arrive at a Fock representation of $\mathcal{A}_{(F)}$. In the quantum theory, each of these solutions $F(\eta,\vx)$ defines a 1-particle state. Again, while the $F(\eta,\vx)$ diverge as `wave functions', they are well-defined as elements of the 1-particle Hilbert space\, $\H$\, because their norm remains (\ref{norm}) finite. (This is somewhat analogous to the fact that in quantum mechanics on $\mathbb{R}^3$, the wave function $\psi(\vec{r}) = e^{-r}/r$ diverges at $r=0$ but provides a well-defined element of the Hilbert space $L^2(\mathbb{R}^3)$.) Since $\hphi(x)$ of (\ref{hphi}) is constructed from the mode functions, as in the $K=0$ case discussed in \cite{ADLS2021}, we will find that it continues to be a well-defined OVD across the big-bang.  

Finally, for $K>0$ the situation is completely analogous. Minor technical differences arise because $k$ is now quantized, and the range of $\ell$ is restricted: (\ref{hphi}) is now replaced by:
\be \label{hphi2} \hphi(\eta, \vx) =  \f{\sqrt{K}}{a(\eta)}\, \sum_{k= \sqrt{K}}^{\infty}  \sum_{\ell=0}^{(k/\sqrt{K})\, -1}\, \sum_{m=-\ell}^{\ell} \Big[ e_k(\eta)\, \mathcal{Y}^{(K)}_{\vk} (\vx)\, A(\vk) \,+\, \bar{e}_k(\eta)\, \bar{\mathcal{Y}}^{(K)}_{\vk}(\vx)\, A^\dag(\vk)\Big]\, ,\ee
and the creation and annihilation operators satisfy now  $[ \h{A}(\vk), \h{A}^\dag(\vk')]=  \f{1}{\sqrt{K}}\, \delta_{ k,  k'}\delta_{\ell, \ell'}\delta_{m,m'}$.\\

 \emph{Remark:} A succinct discussion of the definition and properties of distributions ${\underline{\eta}}^{-n}$ can be found in Appendix A of \cite{ADLS2021}. As noted there, the action of ${\underline{\eta}}^{-1}$ on a test function $f$ is given by the Cauchy principal value:\\
\centerline{$\underline{\eta}^{-1}: f(\eta)\quad \to \quad \lim_{\epsilon\to 0^+} \int_{\mathbb{R}\setminus [-\epsilon, \epsilon]} \rmd \eta\, {\eta}^{-1}\, f(\eta)\,\, =\,\, \int_{0}^\infty \rmd\eta\,\,\, {\eta}^{-1}\, \big(f(\eta) - f(-\eta)\big)$}\\
 which has the intuitively expected property of vanishing for test functions $f(\eta)$ that are even in $\eta$. Distributions ${\underline{\eta}}^{-n}$ are defined by a natural generalization of this procedure and they satisfy (\ref{distribution}). Since our primary goal is to show that there \emph{exists} a consistent extension of QFT across the big-bang, it suffices to make one concrete choice. For concreteness, we will make this choice. 

However, as pointed out in the Appendix A of \cite{ADLS2021}, if one wishes to regard the distribution $\boldsymbol{\eta^{-n}}$ as the algebraic inverse of the function $\eta^n$, then there is an $n$-parameter family of ambiguities in the definition \cite{schwartz,hormander,loja}: $\boldsymbol{\eta^{-n}}  = \underline{\eta}^{-n} + \sum_{i=0}^{n-1} c_i \delta^i(\eta)$, where $c_i$ are constants and $\delta^i$ denotes the $i$th derivative of the Dirac delta distribution. In any case, the Hilbert space structure of 1-particle states defined by $F$  is \emph{insensitive} to this ambiguity (because the volume element vanishes sufficiently rapidly at $\eta=0$). Similarly $\phisq$ is also free of this ambiguity. On the other hand the ambiguity persists in the definition of $\stress$  because, as we shall see in Sec. III and IV, the expression involves $1/a^8(\eta)$ for the radiation filled universe and $1/a^6(\eta)$ for the dust-filled case, while the volume element goes to zero only as $a^4(\eta)$. Use of $\underline\eta^{-n}$ corresponds to fixing this ambiguity by setting $c_i=0$. From a mathematical consideration, this choice results in intuitively expected properties; for $n=1$, for example, it is only when $c_0 =0$ that the distribution sends all even test functions $f(\eta)$ to zero. But it would be much more satisfactory to remove it using physical requirements. This issue is open.

%%%%%%%%%%%%%%%%
\section{Radiation-filled, closed and open FLRW universes }
\label{s3}
%%%%%%%%%%%%%%%%

This section is divided in three parts. In the first, we show that there is a quasi-free representation in which  the OVD $\hphi(x)$ of Eq.~(\ref{hphi}) remains well-defined even when the FLRW space-time is extended across the bing-bang. In the second we focus on the $K<0$ radiation-filled universes and calculate the observables $\phisq$ and $\stress$ using the Hadamard renormalization procedure (summarized in Appendix A), and in the third we carry out these calculations for $K>0$ universes. As expected these results reduce to those in the $K=0$ case reported in \cite{ADLS2021}. 

\subsection{Extension across the big-bang}
\label{s3.1}

We will now apply this general strategy to radiation-filled universes. It is convenient to introduce a variable $L$ with dimensions of length, given by 
\be K = 1/L^2 \quad{\hbox{if $K >0$}} \qquad {\rm and} \qquad K = -1/L^2 \quad{\hbox{if $K <0$}}\,; \ee
the limit $K\to 0$ corresponds to $L\to \infty$ in both cases. For radiation-filled universes, the scale factor is given by 
\be \label{scalefactor1}
a(\eta) = (a_1 L) \,\, \sin (\eta/L),\,\,\, {\hbox{\rm for $K >0$\qquad and}} \qquad  a(\eta) = (a_1 L) \,\sinh (\eta/L),\,\,\, {\hbox{\rm for $K <0\,$}}
\ee
where $a_1$ is a constant (with dimensions  $({\rm length})^{-1}$). The scale factor vanishes at $\eta=0$ and the curvature diverges there. This is the big-bang singularity (and can also be the big-crunch in the $K >0$ case). In the standard analysis one restricts  the range of $\eta$ to lie in $(0, \pi L)$ for closed universes and in $(0, \infty)$ for open universes, so that $\eta=0$ corresponds to the big-bang in both cases. (In the limit $K\to 0$ we have $a(\eta) = a_1\eta$,\, as in the radiation-filled case discussed in \cite{ADLS2021}.)

Since the stress-energy tensor is trace-free, the space-time scalar curvature $R$ vanishes by Einstein's equations. Hence (\ref{timeeq}) becomes $\psi_k^{\prime\prime} + k^2 \psi_k =0$, which can be solved trivially to obtain the general solution
\be
\psi_k(\eta) \,=\, C_k e^{-ik\eta}+D_k e^{ik\eta}
\ee
for both closed and open universes, where $C_k$ and $D_k$ are arbitrary constants. This form suggests that $e_k(\eta) :=\f{1}{\sqrt{2k}}\, e^{-ik\eta}$ would serve as the positive frequency basis satisfying the normalization condition (\ref{normalization}). This expectation is correct: One can systematically arrive at this choice using the following considerations. 

Recall first that the FLRW metric (\ref{metric}) is of the form $g_{ab} = a^2(\eta)\, \go_{ab}$, where $\go_{ab}$ is ultra-static, \, $\go_{ab}\,\rmd x^a\, \rmd x^b \,=\, -\rmd\eta^2 + h_{ij}\, \rmd x^i \rmd x^j$,\, and the spatial metric $h_{ij}$ is of constant curvature.  This ultra-static metric is obviously well-defined if the underlying manifold $M$ is extended across the big-bang, so that $\eta \in (-\infty, \infty)$ in the $K<0$ case and $\eta \in (-\pi L,\, \pi L)$ in the $K>0$ case. Let us denote the extended space-time by $(\Mo, \go_{ab})$. The next step is to note the relation between Klein Gordon equations w.r.t. the physical metric $g_{ab}$ and w.r.t. the ultra-static metric $\go_{ab}$. For \emph{any two} conformally related metrics $g_{ab} = a^2(\eta)\, \go_{ab}$, we have the identity 
\be (\Box -\f{1}{6} R) \phi(x) = a^{-3}(\eta) \,\, (\Boxo - \f{1}{6} \mathring{R})\mathring\phi \qquad {\rm where} \qquad \mathring\phi = a(\eta) \phi\, . \ee 
Since $R=0$ for $g_{ab}$, and $\mathring{R} =6K$, we have 
\be \Box\, \phi(x) = a^{-3}(\eta) \, (\Boxo - K)\mathring\phi\, . \ee
Thus, we are led to solve $(\Boxo - K)\mathring\phi=0$. Using separation of variables, this equation reduces to (\ref{spatialeq}) and (\ref{timeeq}),\, (with $a(\eta)=1, \, \xi=0,\, m^2 = \f{1}{6} R= K$). Therefore we are led to construct the phase space $\pso$ using solutions of the form (\ref{F})\,\, (with $a(\eta) =1$). Thanks to the ultra-static property of $\go_{ab}$,\, $\pso$ admits a unique complex structure $\mathring{J}$. The corresponding quasi-free vacuum is a Hadamard state. In the resulting Fock representation, $\hphio(x)$ is a well-defined OVD on full $(\Mo, \go_{ab})$, given by (\ref{hphi}) (again with $a(\eta)=1$).

Let us return to the the physical FLRW space-time $(M, g_{ab})$ and extend the metric $g_{ab}$ to $\Mo$ simply by letting $\eta$ assume negative values in $a(\eta)$ and, for notational simplicity, continue to denote the extended metric again by $g_{ab}$. (Since $a(\eta)$ is $C^\infty$ on entire $\Mo$, so is the tensor field $g_{ab}$.) It is immediate from the expression (\ref{scalefactor1}) of $a(\eta)$ that $\eta/a(\eta)$ is also a smooth function on $\Mo$. Therefore, it follows that, for every $\Fo(x) \in \pso$,\, 
\be F(x) := \f{1}{a(\eta)}\, \Fo(x)\, \equiv\, \f{1}{\eta}\, \big[\f{\eta}{a(\eta)}\, \Fo(x)\big] \ee
is a well-defined distribution on the extended FLRW space-time $(\Mo, g_{ab})$, satisfying $\Box F(x) =0$. Our phase space $\ps$ will consist of these solutions (with $\Fo(x) \in \pso$). The complex structure on $\pso$ induces a natural complex structure $J$ on $\ps$, with a positive frequency basis $e_k(\eta) :=\f{1}{\sqrt{2k}}\, e^{-ik\eta}$, as anticipated. For reasons discussed in in section \ref{s2.3}, although the solutions $F(x)$ diverge at $\eta=0$ as functions, the norm (\ref{norm}) of the state each $F(x)$ defines in the 1-particle Hilbert space is finite  (and insensitive to the 1-parameter ambiguity in the definition of the distribution $\boldsymbol{\eta^{-1}}$). Finally, $\hphi(x) = \hphio(x)/a(\eta)$ provides us a well-defined OVD on the resulting Fock space satisfying the Klein-Gordon equation in a distributional sense on the extended FLRW space-time $(\Mo, g_{ab})$.\\

\emph{Remarks:}\\
1. While we are considering massless scalar fields and the scalar curvature of the radiation-filled FLRW vanishes, as we see from (\ref{covariance}), $F(x) \not=\Fo(x)$; there is conformal covariance but not invariance. Had we been considering the Maxwell field, we would have had conformal \emph{invariance} --$F_{ab}(x) = \Fo_{ab}(x)$-- and then Fock vacuum in the ultra-static space-time would be the Fock-vacuum in the FLRW space-time. The presence of $1/a(\eta)$ in the relation $F(x) = (1/a(\eta))\, \Fo(x)$ makes the quantum theory of $\hphi(x)$ different from that of $\hphio(x)$.

\noindent 2. In particular, since $(\Mo, \go_{ab})$ is a smooth, globally hyperbolic space-time, the retarded, advanced and the commutator Green's function $\mathring{\Delta} (x, x^\prime) := (\mathring{G}_{\rm Ad} - \mathring{G}_{\rm Ret})(x,\,x^\prime)$ are all well-defined. On the other hand $(\Mo, g_{ab})$ is not globally hyperbolic because of the singularity at $\eta=0$. Therefore, a priori, the retarded and advanced Green's functions are not well-defined unless both the points $x,x^\prime$ lie on the same side of the singularity. Nonetheless, since $\hphi(x) = \hphio(x)/a(\eta)$, the commutator is given by
\ba \label{commutator} 
[\hphi(f),\, \hphi(g)]\,\,  &=:& \,\, i\hbar\,\int_{Mo}\!\! \rmd^4 V\, \int_{Mo}\!\!\rmd^4 V^\prime\, f(x) g(x^\prime) \, \Delta(x,\,x^\prime)\, \h{I} \nonumber\\
&=& \,\, i\hbar \,\int_{Mo} \!\!\rmd^4 x\, \int_{Mo}\!\!\rmd^4 x^\prime\, f(x)a^3(\eta)\,\,\, g(x^\prime) a^3(\eta^\prime)\,\,\, \mathring\Delta(x,\,x^\prime)\,\, \h{I} \ea
and the right hand side is a well-defined for all test functions $f$ and $g$ on the extended space-time $(\Mo, g_{ab})$. The physical correctness of the commutator is assured by following considerations. First, $\Delta(x,\,x^\prime)$ satisfies the Klein-Gordon equation w.r.t. $g_{ab}$ in both arguments. Second, given any $C_0^\infty$  test-function $f$ on $\Mo$, we obtain a solution $F(x)$ by smearing $\Delta(x,\,x^\prime)$ with $f(x^\prime)$. These solutions have smooth Cauchy data away from the $\eta=0$ surface  and are well-defined as distributions on all of $\Mo$. Finally, the commutator (\ref{commutator}) is equivalent to the commutator $[\hPhi(F_1),\, \hPhi(F_2)] = i\hbar\, \Omega(F_1, F_2)\, \h{I}$ on the Fock space that correctly captures the Poisson bracket relations on $\ps$.

\subsection{Observables: $K <0$}
\label{s3.2}

While $\hphi(x)$ is a dimension 1 operator, $\phisq$ has dimension 2 and $\stress$, dimension 4. Therefore, it is not a priori clear whether $\phisq$ and $\stress$ also remain well-defined as distributions in the extended space-time $(\Mo, g_{ab})$. In this sub-section we will compute these quantities and show that they are. To our knowledge these expressions have not appeared in the literature.

Recall that on the fiducial, ultra-static space-time $(\Mo, \go_{ab})$  the vacuum is a Hadamard state: the bi-distribution $\biphio$ it defines has the Hadamard singularity structure as the points are brought together. 
Therefore one might expect that $\biphi = (1/a(\eta)a(\eta^\prime))\,\, \biphio$, would also have the Hadamard ultraviolet behavior away from $\eta=0$ \cite{tadaki}. This expectation is borne out in the following sense. The FLRW metric is spatially homogeneous and isotropic and our choice of mode functions --and therefore of the Fock vacuum they define-- respects these symmetries. Consequently, \emph{as far as spatial directions are concerned,} it suffices to verify whether the $\biphi$ has the Hadamard structure if the points are separated on an $\eta= \eta_0$ surface along any one direction. We verified that this is the case if the separation is in the radial direction and $\eta_0\not=0$. We also verified the Hadamard behavior for points with `purely time-like' separation, i.e., for points of the type $(\vx, \eta)$ and $(\vx, \eta+\epsilon)$ assuming that neither of these two points lie on the $\eta=0$ surface. Note that the result holds also if the two points lie on the opposite sides of the $\eta=0$ surface, i.e., are separated by the big-bang singularity. (Since the metric $g_{ab}$ vanishes at $\eta=0$, the notion of Hadamard behavior is not meaningful if they lies on this surface.)

As explained in Appendix \ref{a1}, in the calculation of $\stress$ it is most convenient to bring together points that have a `purely time-like' separation. Therefore, we will calculate $\biphi$ for two points $x,\, x^\prime$ with $\vx =\vx^\prime$.  
Using Eq. (\ref{completeness}), one obtains: 
\be
\biphi = \frac{\hbar}{4\pi^2 a(\eta)a(\eta-\epsilon)}\int_0^{\infty} dk \,k\, e^{-i k \epsilon}=\lim_{\rm{Im}\, \epsilon \to 0^-}-\frac{\hbar}{4\pi^2 \epsilon^2a(\eta)a(\eta-\epsilon)}
\ee
where $\epsilon \!=\! \eta \!-\!\eta'$. When expanding in $\epsilon<<1$, it reproduces the Hadamard singular structure.  
 
The physical observables, $\biphi$ and $\stress$ can be computed from this bi-distribution. The procedure using Hadamard renormalization (summarized in Appendix~\ref{a1}) is straightforward, but calculations are rather tedious. Using \texttt{Mathematica} and {\it xAct} we obtain:
\be \label{phisq1}
\phisq \,=\, - \frac{\hbar}{48\pi^2 a^2(\eta) L^2}
\ee
and
\bea \label{rhop1}
\left< \h\rho \right>_{\rm ren} & = & \hbar \frac{24 \cosh \left(2 \eta/L\right)\,-9 \,-11 \cosh \left(4 \eta/L\right)}{3840 \pi^{2}\, a^8(\eta)}\,\, a_{1}^{4}\,\,  \label{rho1}\\
\left< \h{p} \right>_{\rm ren} & = & \hbar \frac{-16 \cosh \left(2 \eta/L \right)\,+47 \,+11\, \cosh (4 \eta/L)}{3840 \pi^{2} a^8(\eta)}\,\,  a_{1}^{4}\,\,   \label{p1}
\eea
As a non-trivial check, one can verify that the stress-energy tensor is conserved, i.e., the only non-trivial component $\left< \rho \right>^\prime_{\rm ren}+3\frac{a'}{a}(\left< \rho \right>_{\rm ren}+ \left< p \right>_{\rm ren})=0$ of 
$\nabla^a\stress =0$ is satisfied. Eqs.~(\ref{phisq1}) and (\ref{rhop1}) have two interesting properties:\\
\noindent 
(i) All three expressions are manifestly IR finite; we did not have to introduce an IR regulator;\\
(ii) Away from the singularity, the expressions are manifestly smooth. But they diverge as $\eta^{-n}$ for various values of $n$. Therefore, all these observables are well-defined as distributions on the entire extended space-time $(\Mo, g_{ab})$. Note also that the 4-dimensional volume element goes to zero as $a(\eta)^4$, lowering the effective value of $n$ in the computation of the smeared observables, such as $\int \,\rmd^4V\, f(x) \phisq $. 

The  full\, $\stress$\, can be written in a covariant form using curvature tensors:
\bea
\stress &= & \frac{1}{2880 \pi^{2}} R^{c d} R_{a c b d}-\frac{1}{288 \pi^{2}} \frac{1}{(a L)^{2}}\left(R_{ab}+2R_{acbd}u^c u^d\right) -\frac{11}{1440 \pi^{2}} \frac{1}{(a L)^{4}}\left(g_{a b}+4 u_{a} u_{b}\right)\nonumber\\ \label{knegativeresult}
\eea
where $u^a$ is the unit normal to the $\eta={\rm const}$ surfaces. In the spatially-flat limit $L\to \infty$ we recover the standard result in the literature for a massless minimally coupled scalar field in a spatially flat, radiated-dominated universe \cite{BD}.

\subsection{Observables: $K >0$}
\label{s3.3}

The situation in the $K >0$ case is analogous.  The main differences are just \\
(i) Now $\eta \in (-\pi L,\, \pi L)$ on the extended space-time $(\Mo, g_{ab})$; and,\\
(ii) the integral over $k$ is now a discrete sum and the sum over $\ell$ is bounded above as in (\ref{hphi2}).\\ These differences arise because the global structure of space-time is quite different in the $K>0$ case from that in the $K<0$ case. The vacuum state and the associated modes sense this difference. Therefore, while in our discussion of the two scale factors (\ref{scalefactor1}) one could pass from the $K<0$ to the $K>0$ by replacing the hyperbolic trigonometric functions with trigonometric functions, this simple strategy is no longer valid for the renormalized observables. Therefore, the form of the main equations is quite different from those in Sec.~\ref{s3.2}. 

As in Sec.~\ref{s3.2}, one can compute $\biphi$ using the completeness relation (\ref{completeness}) and it has Hadamard structure in the same sense. For reasons explained there, we can again restrict ourselves to points that are `purely time-like separated'. Then the bi-distribution reads:
\ba \label{biphi2}
\biphi &=&\frac{\hbar}{4\pi^2 L^2a(\eta)a(\eta-\epsilon)}\,\, \sum_{k'=1}^{\infty} k' e^{-i k' \epsilon/L}\nonumber\\
&=&\lim_{\rm{Im}\, \epsilon \to 0^-} \frac{\hbar}{8 \pi^{2}L^2} \frac{1}{a(\eta) a(\eta-\epsilon)} \frac{1}{\left(\cos( \epsilon / L)-1\right)^{2} }
\ea
where we have set $k' = k \, L$ and $\epsilon =\eta-\eta'$. Using \texttt{Mathematica} and {\it xAct} we find (see Appendix A for details):
\be
\phisq =0\, ,
\ee
and
\be \label{rhop2}
\left<\h\rho \right>_{\rm ren}  = \frac{ \hbar\, a_{1}^{4}}{960 \pi^{2} a(\eta)^{8}}  \qquad {\rm and} \qquad
\left< \h{p} \right>_{\rm ren}  =  \frac{\hbar \, a_{1}^{4}}{576 \pi^{2} a(\eta)^{8}}  
\ee
As a non-trivial check, one can verify that the stress-energy tensor is conserved, $\nabla^a \stress=0$, i.e., $\left< \rho \right>^\prime_{\rm ren}+3\frac{a'}{a}(\left< \rho \right>_{\rm ren}+ \left< p \right>_{\rm ren})=0$. Eq.~(\ref{rhop2}) can be written in a manifestly covariant way as
\be
\stress =\frac{1}{2880 \pi^{2}}\, R^{c d} R_{a c b d}
\ee
The three expressions are considerably simpler than those in the $K<0$ case: as noted above one cannot read them off from the $K<0$ expressions because the field modes are  sensitive to the global structure of the spacetime --in particular the topology.\\

\emph{Remark:} At first it may seem surprising that $\phisq =0$ in the $K =0$ \cite{BD}, and $K>0$ cases but not in the $K<0$ case of Sec.~\ref{s3.2}. This difference can be understood in terms of differences in the corresponding classical geometries as follows. Note first that by dimensional considerations, $\phisq$ has the form $\phisq = \hbar S$ where $S$ is a geometric scalar with dimensions $({\rm length})^{-2}$. In radiation-filled universes, there are only three independent candidates: the spatial curvature ${}^3\!R$;\, square of the trace of extrinsic curvature, $(a^{\prime\prime}/a^2)^2$;\, and the time-time component of the 4-dimensional Ricci tensor, $R_{ab}u^au^b$ where $u^a$ is the unit normal to the $\eta= {\rm const}$ slices. It is convenient to write the right side of (\ref{phisq1}) as a linear combination of these three scalars: $\phisq = (1/576)\big(R_{ab} u^a u^b + 6 (a^\prime/a^2)^2 - {}^3\!R\big)$ for $K<0$. In the case when $K=0$ and $K= + 1/L^2$ cases, this combination happens to vanish and we recover the result \,$\phisq =0$\, in these two cases.

%%%%%%%%%%%%%%%%
\section{Dust-filled, closed and open FLRW universes }
\label{s4}
%%%%%%%%%%%%%%%%

In dust-filled FLRW models, the spatial eigenvalue equation (\ref{spatialeq}) is the same as in the radiation-filled case but the time-evolution equation (\ref{timeeq}) is more complicated because the space-time scalar curvature $R$ is now non-zero. As in Sec.~\ref{s3} we will divide our discussion in three parts. In the first we will discuss the extension of the theory across the big-bang; in the second we will discuss observables in the $K<0$ models and, in the third, the $K>0$ models.

\subsection{Extension across the big-Bang}
\label{s4.1}

As in Sec.~\ref{s3.1}, it is convenient to introduce a variable $L$ with dimensions of length, given by $K = 1/L^2$ if $K >0$ and $K = -1/L^2$ if $K <0$; again, the limit $K\to 0$ corresponds to $L\to \infty$ in both cases. For dust-filled universes, the scale factor is then given by 
\ba \label{scalefactor2}
a(\eta) &=& ({-2} a_2 L^2) \,\, (\cos(\eta/L) -1), \quad {\hbox{\rm for $K >0$\qquad and}} \nonumber\\
a(\eta) &=& ({2} a_2 L^2) \,(\cosh(\eta/L) -1), \quad {\hbox{\rm for $K <0\,$}}
\ea
where $a_2$ is a constant with dimensions of $({\rm length})^{-2}$. The scale factor again vanishes at $\eta=0$ and the curvature diverges there. In standard treatments, the range of $\eta$ is restricted to lie in $(0, \pi)$ for closed universes and to lie in $(0, \infty)$ for open universes, so that $\eta=0$ corresponds to the big-bang in both cases. (In the limit $K\to 0$ we have $a(\eta) = a_2\eta^2$,\, as in the dust-filled case discussed in \cite{ADLS2021}.) The space-time metric has the form $g_{ab} = a^2(\eta) \go_{ab}$, where $\go_{ab}$ is the same ultra-static metric that we had in Sec.~\ref{s3} and hence it can be extended to the manifold $\Mo$ on which $\eta$ runs over the full real line, $(-\infty,\, \infty)$. However, with $a(\eta)$ of (\ref{scalefactor2}), the scalar curvature $R$ of $g_{ab}$ does not vanish. Therefore the relation between $\Box$ and $\Boxo$ is more complicated than in Sec.~\ref{s3}. Using the conformal covariance property
\be \label{covariance} \big(\Box - \f{1}{6}R\big)\, F(x) \, =\, a^{-3}(\eta)\, \big(\Boxo - \f{1}{6}\mathring{R}\big)\,\Fo(x) \qquad {\rm with}\quad \Fo(x) = a(\eta)\, F(x)\ee
one finds 
\be \Box F(x) \,=\, 0 \qquad {\rm iff}\qquad \big(\Boxo + \f{a^{\prime\prime}}{a} (\eta)\big)\, \Fo(x)\,=\,0\, .\ee
Thus, now $\Fo$ satisfies the Klein-Gordon equation with a time dependent potential $V(\eta) = \f{a^{\prime\prime}}{a} (\eta)$ on the ultra-static space-time $(\Mo, \go_{ab})$. This makes the analysis more complicated than that for radiation-filled universes. However, as in the $K=0$ case, properties of this potential enable one to select a preferred complex structure on the covariant phase space $\pso$ of solutions $\Fo(x)$ which can then be naturally transferred to the phase space $\ps$ of solutions $F(x)$ to the Klein-Gordon equation on the FLRW space-time $(\Mo, g_{ab})$, using the relation $F(x) = \Fo(x)/a(\eta)$  
(see Sec.~IV of \cite{ADLS2021}). The final results can be summarized as follows.

Let us first consider the $K<0$ case. In this case, the `positive-frequency' basis functions which make the action of the complex structure explicit --as in (\ref{J})-- are given by
\be \label{solutionhyperbolic}
e_k(\eta)= \frac{2kL}{\sqrt{1+4k^2L^2}}\,\Big(1-i \frac{\coth\left(\frac{\eta}{2L}\right)}{2 k L}\Big)\,  \frac{e^{-i k \eta}}{\sqrt{2k}}\, .  
\ee
They satisfy the evolution equation 
\be \label{equationhyperbolic} e^{\prime\prime}_k(\eta) + \Big[k^2 - \f{1}{L^2\big(\cosh(\eta/L) - 1\big)}\, \Big] e_k =0, \ee
the normalization condition (\ref{normalization}), and tend to the basis found in \cite{ADLS2021} in the limit $K \to 0$.  Note that, in contrast to the mode functions $e^{-ik\eta}/\sqrt{2k}$ in the radiation-filled case, each $e_k(\eta)$ now diverges at $\eta=0$ as a function, but $\eta\, e_k(\eta)$ is smooth on entire $\Mo$. Therefore, each $e_k(\eta)$ is well-defined as a distribution on the extended manifold $\Mo$. On the extended FLRW space-time $(\Mo, g_{ab})$, these $e_k(\eta)$ and the spatial basis functions $\mathcal{Y}^{(K)}_{\vk}(\vx)$ can be used to define elements $F(x)$ of $\ps$ as in (\ref{F}), and the OVD $\hphi(x)$ as in (\ref{hphi}). They satisfy the Klein-Gordon equation in a distributional sense.\\

Finally, let us consider the case $K >0$. Now the `positive-frequency' modes satisfying the normalization condition (\ref{normalization}) are given by
\be \label{solutionspheric}
e_k(\eta)= \frac{2kL}{\sqrt{1-4k^2L^2}}\,\Big(1-i \frac{\cot\left(\frac{\eta}{2L}\right)}{2 k L}\Big)\,  \frac{e^{-i k \eta}}{\sqrt{2k}}\,   
\ee
and they satisfy the evolution equation
\be e^{\prime\prime}_k(\eta) + \Big[k^2 + \f{1}{L^2\big(\cos(\eta/L) -1\big)}\, \Big] e_k =0\, . \ee
(Note that these modes and the equation they satisfy can be obtained from (\ref{solutionhyperbolic}) and (\ref{equationhyperbolic}) by the change $L \to i L$. Also, because of the term $kL$ in the numerator --which is absent in the $K=0$ case-- there will be no infrared divergences in the present case.) On the original FLRW space-time $(M, g_{ab})$, $\eta$ takes values in the open interval $(0, \, \pi L)$ with big-bang at $\eta=0$ and big-crunch at $\eta=\pi$. For the extension $\Mo$ of this space-time, we are led to allow $\eta$ to take values in the open interval $(-\pi L,\,\pi L)$ that includes the big-bang of the original space-time (which is now also the big-crunch of the portion of $\Mo$ representing the past big-bang branch). With these changes, the structure is completely analogous to that in the $K<0$ model. 
Therefore, in the $K <0$ equations, we only have to replace the integral over $k$  by a sum (divided by $L$) and restrict the sum over $\ell$ as in (\ref{hphi2}). $F(x)$\, and \,$\hphi(x)$\, are again well-defined and satisfy the Klein-Gordon equation as distributions over the extended FLRW space-time $(\Mo, g_{ab})$.
\subsection{Observables: the $K <0$ case}
\label{s4.2}

Using the expression (\ref{hphi}) of the OVD $\hphi(x)$ in terms of creation and annihilation operators, one can calculate the bi-distribution $\biphi$. It is Hadamard in the same sense as in Sec.~\ref{s3}. More precisely, the mode functions --and hence the Fock vacuum-- are again invariant under the space-time isometries implementing homogeneity and isotropy and we verified that $\biphi$ has the Hadamard structure for nearby points $x,x^\prime$ that have either `purely radial' or `purely time-like' separation (assuming neither of them lies on the $\eta=0$ surface). Again, for reasons explained in Appendix~\ref{a1}, let us focus on points that have `purely time-like' separation. Using \texttt{Mathematica} one obtains: \goodbreak
\bea
\biphi &=& \lim_{\rm{Im}\, \epsilon \to 0^-} \frac{\hbar}{64 L^{2} \pi^{2} a(\eta) a(-\epsilon+\eta)}\,  
\Big[ \Big(3-\cosh \big[\frac{\epsilon}{L}\big]+\cosh \big[\frac{\epsilon-\eta}{L}\big]+\cosh \big[\frac{\eta}{L}\big]\Big)\,\times \nonumber\\
&&\operatorname{Ci}\big[\frac{i \epsilon}{2 L}\big] \operatorname{csch}\big[\frac{\epsilon-\eta}{2 L}\big] \operatorname{csch}\big[\frac{\eta}{2 L}\big] + \frac{1}{\sqrt{\pi}}\, \Big({2 G_{1,3}^{3,1}\big(-\frac{\epsilon ^2}{16 \text{L}^2}|\Scale[0.7]{\begin{array}{c} -1 \\ -1,0,\frac{1}{2} \\ \end{array} }\big) \Big)} \nonumber\\
&+&\frac {1}{\epsilon}\, {2 \sinh \big[\frac{\epsilon}{2 L}\big]\big(I \pi \epsilon-4 L \operatorname{csch}\big[\frac{\epsilon-\eta}{2 L}\big] \operatorname{csch}\big[\frac{\eta}{2 L}\big]+2 \epsilon \operatorname{Shi}\left[\frac{\epsilon}{2 L}\right]\big)} \Big]
\eea
where ${\rm Ci}(z):=-\int_z^{\infty}\cos(t)/t dt$ is the cosine integral function, ${\rm Shi}(z):=\int_0^{z}\sinh(t)/t dt$ is the hyperbolic sine integral function, and\, $G_{1,3}^{3,1}$\, is  Meijer's G function.
Using this expression and Hadamard renormalization implemented using \texttt{Mathematica}, we find 
\be
\phisq = \frac{R}{48\pi^2}\left(\frac{5}{6}-\gamma+\log(\sqrt{2}\mu_{-}\, L a(\eta)) \right)\, - \,\frac{1}{48\pi^2 L^2 a(\eta)^2}
\ee
where  as usual $\gamma$ is the Euler-Mascheroni constant, and $\mu_{-}$ is the renormalization scale for the $K<0$ dust-filled FLRW universe. Note that in the limit $R\to 0$, we recover the result of Sec.~\ref{s3.2} for the $K<0$, radiation-free universe. For stress-energy tensor we obtain
\ba \label{rho3}
 \langle \h\rho_{\rm ren} \rangle  & = & \frac{\hbar a_2^2}{240\pi^2 a^6(\eta)}\,\Big[\, {5\big[54 \log \big( \sqrt{2} a(\eta) L\,\mu_{-} \big)\, -\, 54 \gamma +11\big]\,}\nonumber\\
&+& \, \f{a(\eta)}{a_2L^2}\,  \big[90 \log \big( \sqrt{2} a(\eta) L\, \mu_{-}\big)-90 \gamma +17\big]\, -\, 11 \f{a^2(\eta)}{2a_2^2L^4}\, \Big] \ea 
    %%%%%%%%%%%%%%%%%%%%%%%%%%%%%%%%%%%%%%%%%%%%%%%%%%%%%%%%%%%%%%%%%%
and 
\ba \label{p3}
 \langle \h{p}_{\rm ren} \rangle   & = & \frac{\hbar a_2^2}{720\pi^2\,a^6(\eta)}\, \Big[
 15  \big[-54 \log \big(\sqrt{2} a(\eta) L\, \mu_{-}\,\big)\,+\,54 \gamma + 7\big]\nonumber\\
  &-& \f{2a(\eta)}{a_2L^2}  \big[(-90 \log \big( \sqrt{2} a(\eta) L\, \mu_{-} \big)
  +90 \gamma +28\big]\,-\, 11 \,\f{a^2(\eta)}{2a_2^2 L^4}  \Big]\, .
 \ea
 One can verify that the stress-energy tensor is conserved, $\nabla_a T^{ab}=0$, i.e.,  $\left< \rho \right>^\prime_{\rm ren}+3\frac{a'}{a}(\left< \rho \right>_{\rm ren}+ \left< p \right>_{\rm ren})=0$. In contrast to the radiation-filled universe, these expressions exhibit logarithmic runnings with the renormalization scale $\mu_{-}$. This is a consequence of the fact that now the equation governing time evolution has a potential, $V(\eta) = a^{\prime\prime}/a$. The spatially flat limit is also more subtle than it was in Sec.~\ref{s3} because, while there is no infrared divergence for $K<0$, there is one for $K=0$ \cite{fp}. When these subtleties are taken into account, one recovers the $K=0$ results in the literature.  

Finally, we can collect the results (\ref{rho3}) and (\ref{p3}) to express $\stress$ in terms of space-time geometry: 
\bea
\stress &= & \frac{1}{1728 \pi^{2}} \big( -\frac{9}{4}R^2g_{ab} + R R_{ab} \big) +\frac{1}{64\pi^2}\,\big({\log(\sqrt{2}L\,\mu_{-}\, a(\eta))-\gamma} R R_{ab}\big)\nonumber\\
&-&\frac{1}{1728 \pi^{2}} \frac{1}{(a L)^{2}} \Big(\frac{107}{10}R g_{ab}+R_{ab}\Big)
+ \frac{\log(\sqrt{2}L\,\mu_{-}\, a(\eta))-\gamma}{96 \pi^{2}} \frac{1}{(a L)^{2}}\, \Big(\frac{1}{2}R g_{ab}- 5 R_{ab}\Big)  \nonumber\\
&-&\frac{11}{1440 \pi^{2}} \frac{1}{(a L)^{4}}\,\big(g_{a b}+4 u_{a} u_{b}\big)\,. \eea
%
%%%%%%%%%%%%%%%%
\subsection{Observables: $K >0$}
\label{s4.3}
%%%%%%%%%%%%%%%%

As in Sec.~\ref{s3}, to begin with there are only two main differences from the $K<0$ universes of Sec.~\ref{s4.2} \\
(i) Now $\eta \in (-\pi L,\, \pi L)$ on the extended space-time $(\Mo, g_{ab})$; and,\\
(ii) the integral over $k$ is now a discrete sum and the sum over $\ell$ is bounded above as in (\ref{hphi2}).\\ 
However, the global structure of space-time is quite different in the $K>0$ case from that in the $K<0$ case and, as in Sec.~\ref{s3}, vacua associated with these modes sense this difference. Therefore, there are again notable differences in the expressions of the renormalized observables.

As explained in Appendix \ref{a1}, it suffices to calculate $\biphi$ for points with `purely' time-like separation. The result of this computation using \texttt{Mathematica} is
\bea \hskip-0.8cm
\biphi &=&  \lim_{\rm{Im}\, \epsilon \to 0^-} \hbar\,\, \Scale[0.9]{\frac {e^{-\frac{5 i \epsilon }{2 L}} \csc \left(\frac{\eta }{2 L}\right) \csc ^2\left(\frac{\epsilon }{2 L}\right) \csc \left(\frac{\epsilon -\eta }{2 L}\right) }{192 \pi ^2 L^2 {a}(\eta) {a}(\eta -\epsilon  )}}\times \,\, \Big( 4 \Scale[0.8]{\left(-1+e^{\frac{i \epsilon }{L}}\right)^3}  
{}_3F_2  \Scale[0.8]{\left(\frac{1}{2},2,2;1,\frac{5}{2};e^{-\frac{i \epsilon }{L}}\right) }\nonumber \\
&+&3\, e^{\frac{5 i \epsilon }{2 L}} \Scale[0.9]{\Big[-2 \cos \left(\frac{\epsilon -2 \eta}{2 L}\right)+3 \cos \left(\frac{\epsilon }{2 L}\right)-\cos \left(\frac{3 \epsilon }{2 L}\right)-2\sin^2\left(\frac{\epsilon }{L}\right) \tanh ^{-1}\big(e^{-\frac{i \epsilon }{2 L}}\big)\Big]}\Big) 
\eea
where $_3F_2$ is the hypergeometric function. Using Hadamard renormalization we obtain
\bea
\langle\hat\phi(x)^2\rangle = \frac{R}{48\pi^2}\left(-1+\log(\sqrt{8}\,\mu_{+}\, L a(\eta)) \right) 
\eea
Again, in the limit $R\to 0$ we recover the result of Sec.~\ref{s3.3} for radiation-filled universe with $K >0$.
For the renormalized density and pressure we obtain
\ba \label{rho4}
\langle \h\rho\rangle_{\rm ren}  & = & -\frac{\hbar a_2^2}{240\pi^2\,a^6(\eta)}\, \Big( 26+9 \sin (\frac{\eta}{2L})\big(10 \,\log [\sqrt{8}a(\eta) \mu_{+}\,  L]-7 \big) \Big)\\
\label{p4}
\langle \h{p}\rangle_{\rm ren}  & = & \frac{\hbar a_2^2}{240\pi^2\,a^6(\eta)}\, \Big(5\,\,  \big(54 \log (\sqrt{8}a(\eta) \mu_{+}\,  L)-61\big)-24 \sin^{2}\big(\frac{\eta}{2 L}\big) \big(10 \log( \sqrt{8}a(\eta) \mu_{+}\, L )- 12\big)\, \Big)\nonumber\\ \ea
These expressions again satisfy the conservation equation $\left< \rho \right>^\prime_{\rm ren}+3\frac{a'}{a}(\left< \rho \right>_{\rm ren}+ \left< p \right>_{\rm ren})=0$, and we find logarithmic runnings with the renormalization scale $\mu_{+}$. Finally one can again combine (\ref{rho4}) and (\ref{p4}) to express $\stress$ in terms of space-time geometry:
\bea
\stress &= & \frac{1}{1728 \pi^{2}} \left( R^2g_{ab} -39 R^{cd}R_{acbd} \right) +\frac{\log(\sqrt{8}L\,\mu_{+}\, a(\eta))}{512\pi^2} (12 R^{cd} R_{acbd}-R^2g_{ab}) \nonumber\\
&& +\frac{1}{640 \pi^{2}} \frac{1}{(a L)^{2}} \left( R g_{ab}+30R_{ab}\right)+ \frac{\log(\sqrt{8}L\,\mu_{+}\, a(\eta))}{192 \pi^{2}} \frac{1}{(a L)^{2}} \left(R g_{ab}- 10 R_{ab}\right)   \nonumber\\
\eea

Thus, once again, the observables $\phisq$ and $\stress$ are well-defined distributions on the extended FLRW space-time $(\Mo, g_{ab})$, and realized as smooth functions away from the $\eta=0$ surface.

%%%%%%%%%%%%%%%%
\section{Outlook}
\label{s5}
%%%%%%%%%%%%%%%%

In this work we extended the results of \cite{ADLS2021} to spatially closed and open universes and showed that, in sharp contrast to classical test particles, the big-bang and the big-crunch of the FLRW models are 
quite harmless to linear, test quantum fields. In particular, these singularities do not represent an absolute beginning or absolute end of the universe for them. The analysis of \cite{ADLS2021} was restricted to the $K=0$ FLRW space-times and exploited some simplifications that occur because of spatial flatness. Results of this paper show that the tameness of the big-bang (and the big-crunch) experienced by quantum fields was not an artifact of these simplifications; it is robust vis a vis inclusion of spatial curvature. It turns out that a similar analysis can be carried out also for the Schwarzschild singularity \cite{AS2022} which is more complicated for two reasons: (i) unlike FLRW models, space-time is not spatially isotropic; and, (ii) while the Weyl curvature vanishes identically in the FLRW models, it \emph{diverges} at the Schwarzschild singularity. Nonetheless, the analysis shows that this singularity is again tame when probed with test quantum fields. These results hint at the possibility that all physically interesting space-like singularities of classical general relativity are harmless when one uses physically appropriate tools to investigate their nature \cite{ADRS2022}.

A key point in all these investigations is to recognize that, since quantum fields $\hphi(x)$ are operator-valued distributions (OVDs) already in Minkowski space, one cannot expect them to be better behaved in the vicinity of a singularity. In the more physically oriented treatments of quantum field theory on cosmological space-times, one expands out $\hphi(x)$ using basis functions and it is implicitly assumed that they should be regular functions, as in Minkowski space. But  this is not necessary: For $\hphi(x)$ to be a well-defined OVD, it suffices that they are well-defined as distributions and satisfy the field equation in the distributional sense. We found that this is the case for the basis functions normally used for minimally coupled, massless scalar fields in closed and open FLRW models, even when the space-time is extended across singularity. A priori there is an $n$-parameter ambiguity in defining the distribution $\boldsymbol{\eta^{-n}}$ that corresponds to inverse powers of conformal time $\eta$. However, $\hphi(x)$ as well as $\phisq$ are insensitive to this ambiguity because the scale factor vanishes at $\eta=0$ sufficiently rapidly.

It is also often assumed that the expectation values such as $\stress$ should be smooth tensor fields because the quantum corrected  metric is obtained by solving the semi-classical Einstein's equation with the right side given by $\stress$, and the semi-classical metric is expected to be a smooth tensor field. However,  smoothness of $\stress$ is not essential from the quantum field theory perspective. Since $\hphi(x)$ is an OVD and $\biphi$ is a bi-distribution already in Minkowski space, there is no a priori reason to insist that observables such as $\phisq$ and $\stress$ must be well-behaved functions. In particular, semi-classical corrections in tame regions of space-time could be smooth but if one pushes the theory all the way to the singularity of the classical background, the semi-classical solutions may well become distributional. This is what happens, for example, when one includes the back-reaction to the dynamics of evaporating Callen, Giddings, Harvey, Strominger black holes \cite{cghs}, where quantum corrections soften the singularity and make the semi-classical metric $C^0$ but not $C^1$ there \cite{ori,apr}. Indeed, already in classical general relativity, it has been shown that the metric can be extended across physically interesting Cauchy horizons as a field that is only $C^0$ \cite{md1,md2}, so the curvature diverges there and field equations can hold only in a distributional sense. Returning to semi-classical gravity, one would expect this approximation to cease to be physically accurate already in the Planck regime, even before the singularity is reached. But it could well capture some `faithful shadows' of predictions of a full-fledged quantum gravity theory. For example, there are strong indications from Loop quantum gravity that space-time geometry becomes distributional in a specific sense at Planck scale and the smooth Riemannian geometry of classical general relativity arises only on coarse graining \cite{alrev,crbook,ttbook,30years1,abrev}.  Using the form of the volume element, we found that in both  
Radiation-filled and Dust-filled universes, $\rmd^4 V \stress$ goes as $1/\eta^4$. Therefore, although $\stress$ can be made into a well-defined distribution on the extended space-time, a priori there is a 4-parameter ambiguity in its definition. Since our emphasis was on the existence of an extension of the theory beyond the big-bang, we chose to fix  this ambiguity using the simplest prescription that has intuitively expected mathematical properties. An interesting open issue is whether this ambiguity can be removed by imposing compelling physical requirements. Perhaps these will  descend from quantum gravity considerations. More generally, it is of considerable interest to better understand possible relations between the appearance of fields that satisfy the desired equations only in a distributional sense at these three different levels --classical general  relativity, quantum field theory in curved space-time and loop quantum gravity \cite{ADRS2022}.
   
This investigation also suggests some directions for future work in quantum field theory in curved space-times by itself. Chronologically, the theory was developed through the introduction of concrete Fock representations of the canonical commutation relations, by expressing $\hphi(x)$ as a sum of creation and annihilation operators associated with suitable basis functions (or modes) in specific space-times. Through a series of sustained mathematical advances, the theory has been made significantly more general and formulated in generic globally hyperbolic space-times. The emphasis has shifted to an algebraic approach  in which one introduces abstractly defined operator algebras (such as $\mathcal{A}_{(f)}$ and $\W$ of Sec. \ref{s2}), and works with regular states (generally taken to be Hadamard) without necessarily constructing the  associated GNS representation (see, e.g., \cite{bfv,BDH,hw,cfkr}). Powerful techniques from micro-local analysis and wavefront sets have added much rigor and generality to the framework. However, at their core these developments appear to be deeply intertwined with notion of global hyperbolicity, and therefore not applicable --at least directly-- once the space-time is extended to include singularities, such as the $(\Mo, g_{ab})$'s considered in sections \ref{s2} - \ref{s4}. On the other hand, we saw that one \emph{can} extend quantum field theory to such space-times. The strategy outlined in Sec.~\ref{s2.3} shifts the emphasis from the abstract algebra $\mathcal{A}_{(f)}$ generated by $\hphi(f)$ to the algebra $\mathcal{A}_{(F)}$ generated by $\hPhi(F)$ associated with suitable classical solutions $F(x)$ to the field equations, and the corresponding Weyl operators $\h{W}(F)$. In globally hyperbolic space-times, we have well-defined retarded, advanced and commutator distributions, and one can readily go back and forth between $\hphi(f)$ and $\hPhi(F)$ (and hence between the corresponding Weyl operators). But once the FLRW space-time $(M, g_{ab})$ is extended to $(\Mo, g_{ab})$ that includes the big-bang/big-crunch singularity, we no longer have global hyperbolicity and it is not obvious that the powerful techniques that have been developed for globally hyperbolic space-times can be used. Nonetheless, we could associate operators $\hPhi(F)$ with certain judiciously chosen \emph{distributional} classical solutions $F(x)$ for which, in particular, the 1-particle norm (\ref{norm}) is well-defined in spite of the distributional character of the solutions. Of course, space-times considered so far --FLRW models and the Schwarzschild `interior'-- are very special, like the cosmological space-times and Rindler wedges that were the focus of attention in the early days of quantum field theory in curved space-times. But as far as singularities are concerned, the FLRW and the Schwarzschild space-times are among the most interesting examples from a physical perspective. Therefore, it would well-worth investigating whether the powerful mathematical methods that have been introduced over the last 2-3 decades can be extended to a judiciously chosen class of singular space-times. Understanding the structure of quantum fields from a general perspective would provide us with significant new insights on way to full quantum gravity which, we believe will be singularity free because of the ultraviolet regularity of quantum geometry. Perhaps 
distributional geometries provided by an appropriate generalization of the current semi-classical gravity will provide a bridge between the low energy continuum geometries and the full quantum Riemannian geometry at Planck scale, such as the one provided by loop quantum gravity. We hope that results presented here and other related works will open a door to extend the well-developed, mathematically rigorous quantum field theory in curved space-times in these directions which have remained unexplored so far. \bigskip

%%%%%%%%%%%%%%%%
\section*{Acknowledgements}
%%%%%%%%%%%%%%%%

We would like to thank Marc Schneider for numerous discussions as well as comments on the manuscript, and
the York Mathematical Physics group of comments and questions during a seminar on this subject. This work was supported by the NSF grant PHY-1806356, and the Eberly Chair funds of Penn State. We acknowledge extensive use of some packages of {\bf xAct} for \texttt{Mathematica}. 

\appendix

%%%%%%%%%%%%%%%%
\section{Hadamard renormalization}
\label{a1}
%%%%%%%%%%%%%%%%

For convenience of the reader, in this Appendix we call the renormalization procedure that underlies our calculations of $\langle \hat\phi^{2}(x)\rangle$ and $\left\langle T_{a b}(x)\right\rangle$ used in sections~\ref{s3} and \ref{s4}. Given the  bi-distribution $\left<\phi(x)\phi(x')\right>$ it is possible to compute these observables in a straightforward manner using Hadamard renormalization.  We review here the basic formalism, following Ref. \cite{Hadamard}.

As is well-known, the product of two operator-valued distributions evaluated at the same point is not well-defined. The first step towards renormalization typically involves the regularization of divergences by splitting the points. Let $\mathcal U$ denote a convex normal neighborhood of the spacetime manifold $M$. For any two points, $x, x' \in \mathcal U$,  there exists a unique geodesic that connects $x$ and $x'$ and which lies entirely in $\mathcal U$ \cite{Oneil}. Let $\tau(x,x')$ denote the geodesic distance between $x$ and $x'$, and $t^a(x)$ the geodesic's tangent vector at $x$, normalized as $g_{ab}t^a t^b=\epsilon$ (where $\epsilon=\pm 1$ for space-like/time-like geodesics, respectively).  It is convenient to work instead  with $\sigma^a(x,x'):=\tau(x,x')\,t^a(x)$, which is a rescaled tangent vector of the geodesic at $x$, with norm equal to the geodesic distance. This is  a vector at $x$ and a scalar at $x'$.  This bi-tensor naturally leads to the Synge's world function $\sigma(x,x'):=\frac{1}{2}\sigma_a(x,x')\sigma^a(x,x')=\frac{\epsilon}{2}\tau^2(x,x')$. 

For sufficiently close points $x\neq x'$, the  bi-distribution $\left<\phi(x)\phi(x')\right>$ computed in any Hadamard state has the following singular structure:
\footnote{The precise sense in which the quasi-free states used in the main text satisfy Hadamard conditions is spelled out in Secs.~\ref{s3} and \ref{s4}.}
\bea \biphi 
=\frac{1}{8 \pi^{2}}\left\{\frac{U\left(x, x^{\prime}\right)}{\sigma\left(x, x^{\prime}\right)}+V\left(x, x^{\prime}\right) \log \sigma\left(x, x^{\prime}\right)+W\left(x, x^{\prime}\right)\right\}
\eea
where $U\left(x, x^{\prime}\right)$ and $V\left(x, x^{\prime}\right)$ are smooth, real-valued bi-functions that depend only on the local geometry of the spacetime, while $W\left(x, x^{\prime}\right)$ is a smooth, real bi-function that encodes the information about the quantum  state chosen. 

It is customary to fix $x$, interpret $U\left(x, x^{\prime}\right), V\left(x, x^{\prime}\right), W\left(x, x^{\prime}\right)$ as functions of $x'$, and then expand them in covariant Taylor expansions around the point $x$ \cite{Christensen}:
\bea
U\left(x, x^{\prime}\right) & = & u(x)+\sum_{p=1}^{\infty} \frac{(-1)^{p}}{p !} u_{a_{1} \dots a_p}(x) \sigma^{a_{1}}\dots  \sigma^{a_{p}}\\
V\left(x, x^{\prime}\right) & = & \sum_{n=0}^{\infty} V_{n}\left(x, x^{\prime}\right) \sigma^{n}, \, 
\quad V_{n}\left(x, x^{\prime}\right)  =  v_{n}(x)+\sum_{p=1}^{\infty} \frac{(-1)^{p}}{p !} v_{n a_{1} \dots a_p}(x) \sigma^{a_{1}}\dots  \sigma^{a_{p}}\\
W\left(x, x^{\prime}\right) & = & \omega(x)+\sum_{p=1}^{\infty} \frac{(-1)^{p}}{p !} \omega_{a_{1}\dots  a_p}(x) \sigma^{a_{1}}\dots  \sigma^{a_{p}}\, .
\eea
The renormalized quantities of interest are given then by the following formulas, which involve some of the Taylor coefficients in the previous expansions:
\bea
\phisq   
& = & \frac{1}{8 \pi^{2}}\left(\omega(x)-v_{0}(x) \log \mu^{2}\right) \label{O1}\\
\stress 
& = & \frac{1}{8 \pi^{2}}\left\{-\omega_{ a b}(x)+\left[\frac{1}{2}-\xi\right] \nabla_{a} \nabla_{b} \omega(x)+\left[\xi-\frac{1}{4}\right]  g_{a b} \square \omega(x)+\xi R_{a b}\, \omega(x) \right\} \nonumber\\
 & + & \frac{ \log \mu^{2}}{8 \pi^{2}}\left\{v_{0 a b}(x)+g_{a b} v_{1}(x)-\left[\frac{1}{2}-\xi\right] \nabla_{a} \nabla_{b} v_{0}(x)-\left[\xi-\frac{1}{4}\right] g_{a b} \square v_{0}(x)-\xi R_{a b} v_{0}(x)\right\}  \nonumber\\
 & & -\frac{1}{8\pi^2}g_{ab}v_1(x)\, . \label{Thadamard}
\eea
where $\mu>0$ is an arbitrary renormalization scale. Notice that the first line in (\ref{Thadamard}) contains all the information about the  quantum state, while the third line contains the  term that gives the trace anomaly.  Using identities from the Hadamard formalism, it can be proven that $\nabla_a \stress\, =\, 0$. 

The geometric contributions, $v_0(x)$, $v_1(x)$, $v_{0ab}(x)$, are universal (in the sense that do not depend on the choice of the Hadamard state), and are determined by the background metric according to   expressions (\ref{v0}), (\ref{v1}) below. In particular, it is not difficult to see that conformally coupled massless fields do not run with the renormalization scale.  
In contrast, the quantities $\omega(x)$ and $\omega_{0ab}(x)$, which depend on the choice of quantum state, must be computed in detail for each problem. This is the non-trivial part of the calculation.
If we define the singular part of the distribution as
\bea
\langle \hat\phi(x) \hat\phi\left(x^{\prime}\right)\rangle_{\text{sing} }:=\frac{1}{8 \pi^{2}}\left\{\frac{U\left(x, x^{\prime}\right)}{\sigma\left(x, x^{\prime}\right)}+V\left(x, x^{\prime}\right) \log \sigma\left(x, x^{\prime}\right)\right\} \label{singular}
\eea
then, 
$\omega(x)$ and $\omega_{0ab}(x)$ can be obtained from the  bi-distribution by directly applying the formulas:
\bea
W\left(x, x^{\prime}\right)=8 \pi^{2}\left[\left\langle\phi(x) \phi\left(x^{\prime}\right)\right\rangle-\left\langle\phi(x) \phi\left(x^{\prime}\right)\right\rangle_{\operatorname{sing}}\right] \label{W}
\eea
\bea
\omega(x) & = & \lim _{x^{'}\to  x} W\left(x, x^{\prime}\right) \label{w}\\
\omega_{ab}(x) & = & \lim _{x^{\prime} \rightarrow x} \nabla_{a} \nabla_{b} W\left(x, x' \right) \label{wab}
\eea

The strategy to obtain the renormalized observables (\ref{O1}), (\ref{Thadamard}) is the following. We fix two  points $x$ and $x'$ of the spacetime manifold, and compute $\langle\phi(x) \phi\left(x^{\prime}\right)\rangle$ (the strategy for choosing these two points in the most convenient manner is discussed below).  Then, we compute $\langle \hat\phi(x) \hat\phi\left(x^{\prime}\right)\rangle_{\text{sing} }$ using (\ref{singular}). To get this quantity we need to evaluate $U(x,x')$, $V(x,x')$ and $\sigma(x,x')$ to the required order in point-splitting. Up to the required order  to compute (\ref{w}) - (\ref{wab}) in 4 dimensions,  the covariant Taylor expansion of the functions $U(x,x')$ and $V(x,x')$ are
\bea
U(x,x') &  = & u(x)-u_{a}(x) \sigma^{a}+\frac{1}{2 !} u_{a b}(x) \sigma^{ a} \sigma^{ b}-\frac{1}{3 !} u_{a b c}(x) \sigma^{ a} \sigma^{ b} \sigma^{ c} +\frac{1}{4 !} u_{a b c d}(x) \sigma^{ a} \sigma^{b} \sigma^{ c} \sigma^{ d}+O\left(\sigma^{5 / 2}\right) \nonumber\\
V(x,x') & = & V_{0}(x,x')+V_{1}(x,x') \sigma(x,x')+O\left(\sigma^{3 / 2}\right) 
\eea
with 
\bea
V_{0}(x,x') & = & v_{0}(x)-v_{0 a}(x) \sigma^{ a}+\frac{1}{2 !} v_{0 a b}(x) \sigma^{ a} \sigma^{ b}+O\left(\sigma^{3 / 2}\right) \\
V_{1}(x,x') & = & v_{1}(x)+O\left(\sigma^{1 / 2}\right)
\eea
The coefficients in this expansion are completely determined by the background metric, according to
\bea
u_{0} & = & 1, \quad \quad u_{0 a}=0, \quad \quad u_{0 a b}=\frac{1}{6} R_{a b}, \quad \quad u_{0 a b c}=\frac{1}{4} R_{(a b ; c)}, \\
u_{0 a b c d} & = & \frac{3}{10} R_{(a b ; c d)}+\frac{1}{12} R_{(a b} R_{c d)} +\frac{1}{15} R_{p(a|q| b} R_{c d)}^{p q}
\eea
and 
\bea
v_{0} &  = &   \frac{1}{2}\left[ m^{2}+\left(\xi-\frac{1}{6}\right) \right] R,   \quad \quad v_{0a}  =  \frac{1}{4}\left(\xi-\frac{1}{6}\right) R_{; a},  \label{v0}\\
v_{0ab}  & = &  \frac{1}{12} m^{2} R_{a b}+\frac{1}{6}\left(\xi-\frac{3}{20}\right) R_{; a b} -\frac{1}{120} \square R_{a b}+\frac{1}{12} \left(\xi-\frac{1}{6}\right) R R_{a b} \nonumber\\
&& +\frac{1}{90} R_{a}^{p} R_{p b}-\frac{1}{180} R^{p q} R_{p a q b} - \frac{1}{180} R_{a}^{p q r} R_{p q r b} \nonumber
\eea
and 
\begin{equation}
\begin{aligned}
v_{1} &=\frac{m^{4}}{8} +\frac{m^{2} R}{4} \left[\xi-\frac{1}{6}\right]  - \frac{\square R}{24}\left[\xi-\frac{1}{5}\right] +\frac{R^{2}}{8} \left[\xi-\frac{1}{8}\right]^{2}  -\frac{(R_{p q} R^{p q}- R_{p q r s} R^{p q r s})
}{720} 
\end{aligned} \label{v1}
\end{equation}
On the other hand, to obtain the  bi-tensor $\sigma^a(x,x')$ we follow Appendix B of \cite{AHS95}, and from this we can readily obtain the Synge world's function by $\sigma(x,x'):=\frac{1}{2}\sigma_a(x,x')\sigma^a(x,x')$. 

While the procedure just described can be implemented for any two points $x,x'$ in a normal neighborhood, the  specific calculation can be greatly simplified if $x,x'$ are chosen conveniently. For a quantum state that respects the homogeneity and isotropy of the FLRW spacetime background, $\stress$ has the perfect fluid form, $\stress=(\langle \h\rho\rangle_{\rm ren}+\langle \h{p}\rangle)u_au_b\,+\,\langle p\rangle_{\rm ren}\,\, g_{ab}$, with $u^a$, the unit time-like normal to the $\eta={\rm const}$ slices, and 
\bea
\left<\h\rho\right>_{\rm ren} & = & u^a u^b \stress \label{rho}\\
\left<\h{p}\right>_{\rm ren} & = & \frac{1}{3}\, h^{ab}\, \stress = \frac{1}{3} \left<\h\rho\right>_{\rm ren}\, +\, \langle \h{T}_{a}^a\rangle_{\rm ren} \label{p}
\eea
It follows from Eqs.~\ref{Thadamard},\, \ref{w} and \ref{wab} that, to evaluate $\left<\h\rho\right>_{\rm ren}$ we need $\omega(x)$, $\omega_{ab}(x)u^a u^b$ and to evaluate $\langle\h{T}_{a}^a\rangle_{\rm ren}$ we need $\omega_a^a(x)$. Finally, because of the identity $\omega^a_a(x)=(m^2+\xi R)\omega(x)-6v_1(x)$ from the Hadamard formalism, one only needs to calculate $\omega(x)$ and $\omega_{ab}(x)u^a u^b$. To compute $\omega(x)$, one can take the point-splitting in any direction. However, to calculate $\omega_{ab}(x)u^a u^b=\frac{1}{a(\eta)^2}\,\omega_{\eta\eta}(x)$ it is crucial to do the splitting in the time-like direction. Therefore, it suffices to calculate the bi-distribution for points that have a `purely time-like' separation: $x=(\eta,\chi,\theta,\phi)$, $x'=(\eta',\chi,\theta,\phi)$. Then, the quantity of interest $\omega_{\eta\eta}(x)$ emerges in the covariant Taylor expansion:
\bea
W\left(x, x^{\prime}\right)=w(x)-w_{\eta}(x)\left(\eta-\eta^{\prime}\right)+\ldots+\frac{1}{2} w_{\eta \eta}(x)\left(\eta-\eta^{\prime}\right)^{2}+\ldots
\eea

To perform the calculation we express both $\left\langle\phi(x) \phi\left(x^{\prime}\right)\right\rangle$ and $\left\langle\phi(x) \phi\left(x^{\prime}\right)\right\rangle_{\operatorname{sing}}$ in terms of $\epsilon=\eta- \eta'$, compute $\hat W(x,\epsilon)\equiv W(x,x-\epsilon)$ using (\ref{W}), and then we evaluate
\bea
\omega(x) & = & \left. \hat W(x,\epsilon)\right|_{\epsilon=0}\\
\omega_{\eta\eta}(x) & = & a^2 u^au^b\left. \nabla_a \nabla_b \hat W(x,\epsilon)\right|_{\epsilon=0} = \left.\left(\partial_{\eta}^{2}-H \partial_{\eta}\right) \hat W\left(x, \epsilon \right) \right|_{\epsilon=0}\, .
\eea
We then calculate the rights sides of (\ref{rho})-(\ref{p}) using these results and equation (\ref{Thadamard}). The result of doing this calculation for the 4 spacetimes considered in the main text yields:\\   (i) equations (\ref{rho1})-(\ref{p1}) for the radiation filled, $K=-1/L^2$ universe;\,\, (ii) equations (\ref{rhop2}) for the radiation-filled $K=+1/L^2$ universe;\,\, (iii) equations (\ref{rho3})-(\ref{p3}) for the dust-filled, $K=-1/L^2$ universe;\,\, and,\, (iv) equations (\ref{rho4})-(\ref{p4}) for the dust-filled $K=+1/L^2$ universe.

A useful identity that can be used  as a check during intermediate calculations is $\omega_a(x)=\frac{1}{2}\nabla_a \omega(x)$.

%%%%%%%%%%%%%%%%
\section{Radial component of the Klein-Gordon equation} 
\label{a2}
%%%%%%%%%%%%%%%%

For completeness, in this appendix we show how to obtain solutions (\ref{radialmodes}) to the radial equation (\ref{spatialeq}) for the field modes, as well as their main properties and the orthonormality (\ref{ortho1}) and the addition formula (\ref{completeness}). Several of these results are stated without proof in the literature (e.g., \cite{eigenvalues}) but we were not able to find a complete treatment. This Appendix provides an essentially self-contained derivation for the convenience of the reader.

\subsection{Main solutions and properties}

With the ansatz (\ref{decom}) the spatial equation (\ref{radialmodes}) reduces to
\bea
\left[\partial_{\chi}^2+2\sqrt{K}\cot(\sqrt{K}\chi) \partial_{\chi} -\frac{K \ell(\ell+1)}{\sin^2(\sqrt{K}\chi)} +k^2-K\right] \Pi_{K, k\ell}=0 \label{maineq}
\eea
First we note that the solution to this equation is closely related to the associated Legendre polynomials.  
These are solutions to 
\bea
\left(1-z^{2}\right) \frac{d^{2} u}{d z^{2}}-2 z \frac{d u}{d z}+\left[\nu(\nu+1)-\frac{\mu^{2}}{1-z^{2}}\right] u=0\, .
\eea
By performing a change of variables $u=v(z^2-1)^{1/4}$, and $z=\cos(\sqrt{K}x)$, this equation becomes
\bea
\left[\partial_{x}^2+2\sqrt{K}\cot(\sqrt{K}x) \partial_{x} +K\left[ \nu(\nu+1)-\frac{3}{4}-\frac{\mu^2-\frac{1}{4}}{\sin(\sqrt{K}x)} \right] \right] v=0 \, .
\eea
This equation can be identified with (\ref{maineq}) provided that $\mu^2-\frac{1}{4}=\ell(\ell+1)$ and $K(\nu(\nu+1)-\frac{3}{4})=k^2-K$. We can solve these equations to get $\mu=\pm(\ell+\frac{1}{2})$ and $\nu=-\frac{1}{2}\pm \frac{k}{\sqrt{K}}$. Therefore, 
\bea
\frac{1}{\sqrt{\sin(\sqrt{K}x)}}P_{-\frac{1}{2}+ \frac{k}{\sqrt{K}}}^{\pm(\ell+\frac{1}{2})}(\cos(\sqrt{K}x))
\eea
are two linearly independent solutions to (\ref{maineq}). 
\footnote{For $K=1/L^2$, $\mu_++\nu$ is an integer but $\mu_+$ isn't, so by statement 8.707 (4) in \cite{GR} $P^{\mu_+}_{\nu}(z)$ and $P^{-\mu_+}_{\nu}(z)$ are linearly independent. On the other hand, from the identity $P_{-\nu-1}^{\mu}(x)=P_{\nu}^{\mu}(x)$ (see eqn 8.733(5) in \cite{GR}) we get $P_{-\frac{1}{2}+ \frac{k}{\sqrt{K}}}^{\mu}(x)=P_{-\frac{1}{2}- \frac{k}{\sqrt{K}}}^{\mu}(x)$ so the one of the two signs of $\nu_{\pm}$ is redundant.}
We have to show now its connection to  (\ref{radialmodes}). Let us focus first on the $+$ solution. We will use the identity (see eqn 8.733(1) in \cite{GR})
\bea
\left(1-x^{2}\right) \frac{d P_{\nu}^{\mu}(x)}{d x}=-\sqrt{1-x^{2}} P_{\nu}^{\mu+1}(x)-\mu x P_{\nu}^{\mu}(x)
\eea

From this identity one can easily infer
\bea
P^{a+1}_{\nu}(\cos(\sqrt{K}x)) & = & -\sin(\sqrt{K}x) \frac{d}{d(\cos(\sqrt{K}x))} P^{a}_{\nu}(\cos(\sqrt{K}x))-a \frac{\cos(\sqrt{K}x)}{\sin(\sqrt{K}x)} P^{a}_{\nu}(\cos(\sqrt{K}x))\nonumber \\
 & = & \frac{1}{\sqrt{K}}\frac{d}{dx} P^{a}_{\nu}(\cos(\sqrt{K}x))-a \frac{\cos(\sqrt{K}x)}{\sin(\sqrt{K}x)} P^{a}_{\nu}(\cos(\sqrt{K}x))  \nonumber \\
 & = & \frac{\sin^{a}(\sqrt{K}x)}{\sqrt{K}}\frac{d}{dx} \frac{P^{a}_{\nu}(\cos(\sqrt{K}x))}{\sin^{a}(\sqrt{K}x)} \nonumber\\
 & = &  -\sin^{a+1}(\sqrt{K}x)\frac{d}{d\cos(\sqrt{K}x)} \frac{P^{a}_{\nu}(\cos(\sqrt{K}x))}{\sin^{a}(\sqrt{K}x)} \label{auxlogic}
\eea
Proceeding recursively, we obtain:
\bea
P^{a+\ell}_{\nu}(\cos(\sqrt{K}x)) & = & (-1)^{\ell} \sin^{a+\ell}(\sqrt{K}x)  \left[\frac{d}{d\cos(\sqrt{K}x)}\right]^{\ell} \frac{P^{a}_{\nu}(\cos(\sqrt{K}x))}{\sin^{a}(\sqrt{K}x)}
\eea
For $a=1/2$ we have (see eqn 8.754(1) in \cite{GR}), $P^{1/2}_{\nu}(\cos(\sqrt{K}x))=\sqrt{\frac{2}{\pi \sin(\sqrt{K}x)}} \cos(k x)$, so, up to an irrelevant constant, one concludes
\bea
\frac{1}{\sqrt{\sin(\sqrt{K}x)}}P_{-\frac{1}{2}+ \frac{k}{\sqrt{K}}}^{(\ell+\frac{1}{2})}(\cos(\sqrt{K}x)) \sim   \sin^{\ell}(\sqrt{K}x)  \left[\frac{d}{d\cos(\sqrt{K}x)}\right]^{\ell+1} \sin(k x) \label{uninteresting}
\eea
which is one of the two linearly independent solutions of (\ref{radialmodes}).  Let us focus now on the $-$ solution above.  For this we will use the following identity (see eqn 8.731, 1(2) in \cite{GR}):
\bea
\left(z^{2}-1\right) \frac{d P_{\nu}^{\mu}(z)}{d z}=(\nu+\mu)(\nu-\mu+1) \sqrt{z^{2}-1} P_{\nu}^{\mu-1}(z)-\mu z P_{\nu}^{\mu}(z)
\eea
This identities yields
\bea
(\nu+a)(\nu-a+1)\!\!&&\!\!P^{a-1}_{\nu}(\cos(\sqrt{K}x))\nonumber\\ 
& = &  +i\sin(\sqrt{K}x) \frac{d}{d(\cos(\sqrt{K}x))} P^{a}_{\nu}(\cos(\sqrt{K}x))+\frac{a}{i} \frac{\cos(\sqrt{K}x)}{\sin(\sqrt{K}x)} P^{a}_{\nu}(\cos(\sqrt{K}x)) \nonumber\\
& = & -i \sin^{-a+1}(\sqrt{K}x)\frac{d}{d\cos(\sqrt{K}x)} \frac{P^{a}_{\nu}(\cos(\sqrt{K}x))}{\sin^{-a}(\sqrt{K}x)}
\eea
We have skipped most of the steps, as the calculation is very similar to (\ref{auxlogic}). Proceeding recursively:
\bea
   P^{a-\ell}_{\nu}(\cos(\sqrt{K}x)) & = & -i  \frac{\Gamma(\nu-a+1) \Gamma(\nu+a-\ell+1)}{\Gamma(\nu-a+\ell+1)\Gamma(\nu+a+1)}  \sin^{-a+\ell}(\sqrt{K}x) \left[\frac{d}{d\cos(\sqrt{K}x)}\right]^{\ell} \frac{P^{a}_{\nu}(\cos(\sqrt{K}x))}{\sin^{-a}(\sqrt{K}x)}\nonumber
\eea
For $a=-1/2$ we have (see eqn 8.754(3) in \cite{GR}), $P^{-1/2}_{\nu}(\cos(\sqrt{K}x))=\sqrt{\frac{2}{\pi \sin(\sqrt{K}x)}}\frac{\sqrt{K}}{k} \sin(k x)$. As a result, we obtain
\bea
\frac{P_{-\frac{1}{2}+ \frac{k}{\sqrt{K}}}^{-(\ell+\frac{1}{2})}(\cos(\sqrt{K}x))}{\sqrt{\sin(\sqrt{K}x)}} =\frac{\sqrt{2K}}{\sqrt{\pi}k} \frac{ -i\,\Gamma(k/\sqrt{K}-\ell)}{ \Gamma(k/\sqrt{K}+\ell+1)}   \sin^{\ell}(\sqrt{K}x)  \left[\frac{d}{d\cos(\sqrt{K}x)}\right]^{\ell+1} \cos(k x)  \quad \quad \quad \text{} \label{interesting}
\eea
which is the remaining  linearly indepenent solution of (\ref{radialmodes}).

Let us analyze the behavior of these two linearly independent solutions (\ref{uninteresting}) and (\ref{interesting}) when $x=0$. Because of the prefactor $\sin^{\ell}(\sqrt{K}x)$ in both of them, the result vanishes at $x=0$ unless we can extract an equal number of powers of $\sin(\sqrt{K}x)$ from the derivative $\frac{d}{d\cos(\sqrt{K}x)}=\frac{1}{-\sqrt{K}\sin(\sqrt{K}x)}\frac{d}{dx}$. For (\ref{interesting}) we obtain
\bea
\left. \sin^{\ell}(\sqrt{K}x)  \left[\frac{d}{d\cos(\sqrt{K}x)}\right]^{\ell+1} \cos(k x)\right|_{x=0} &=& \left. \frac{1}{(-\sqrt{K})^{\ell} }\frac{d^{\ell}}{dx^{\ell}} \frac{d\cos(k x)}{d\cos(\sqrt{K}x)} \right|_{x=0} \nonumber\\
& =&  \left. \frac{-k}{(-\sqrt{K})^{\ell+1} }\frac{d^{\ell}}{dx^{\ell}} \frac{\sin(k x)}{\sin(\sqrt{K}x)} \right|_{x=0}
\eea
The function $\frac{\sin(k x)}{\sin(\sqrt{K}x)}$ is  smooth  around $x=0$, so its derivatives are all well-defined for any $\ell$ at $x=0$. Let us do a similar analysis for (\ref{uninteresting})
\bea
\left. \sin^{\ell}(\sqrt{K}x)  \left[\frac{d}{d\cos(\sqrt{K}x)}\right]^{\ell+1} \sin(k x)\right|_{x=0} &= & \left. \frac{1}{(-\sqrt{K})^{\ell} }\frac{d^{\ell}}{dx^{\ell}} \frac{d\sin(k x)}{d\cos(\sqrt{K}x)} \right|_{x=0} \nonumber\\
&=&  \left. \frac{k}{(-\sqrt{K})^{\ell+1} }\frac{d^{\ell}}{dx^{\ell}} \frac{\cos(k x)}{\sin(\sqrt{K}x)} \right|_{x=0}
\eea
The function $ \frac{\cos(k x)}{\sin(\sqrt{K}x)}$ and all its derivatives, in contrast to the previous case, are singular at $x=0$. If the general solution of (\ref{maineq}) is a linear combination of (\ref{uninteresting}) and (\ref{interesting}), demanding regularity at the origin $x=0$ requires the coefficient of (\ref{uninteresting}) to be  zero.

For $K=+1/L^2$, the coordinate $x$ is bounded above by $\pi L$, and one can do a similar analysis of regularity around this point. The  prefactor $\sin^{\ell}(\sqrt{K}x)=\sin^{\ell}(x/L)$ again vanishes when $x=\pi L$:
\bea
\left. \sin^{\ell}(\sqrt{K}x)  \left[\frac{d}{d\cos(\sqrt{K}x)}\right]^{\ell+1} \cos(k x)\right|_{x=\pi L}  =  \left. \frac{-k}{(-\sqrt{K})^{\ell+1} }\left[\frac{d}{dx}\right]^{\ell} \frac{\sin(k x)}{\sin(\sqrt{K}x)} \right|_{x=\pi L}
\eea
The function $\frac{\sin(k x)}{\sin(\sqrt{K}x)}$ is not smooth at $x=\pi L$ unless we demand $kL\in \mathbb Z$.
The property $kL\in \mathbb Z$ imposes in turn another constraint. Indeed, from the trigonometric identity 
\bea
\cos (n x) =\sum_{i=0}^{n / 2} \sum_{j=0}^{i}(-1)^{i-j}\left(\begin{array}{c}
n \\
2 i
\end{array}\right)\left(\begin{array}{l}
i \\
j
\end{array}\right) \cos ^{n-2(i-j)} x \label{trigoid}
\eea
valid when $n$ is integer, one infers that $\cos(kx)$ is a polynomial of $\cos(\sqrt{K}x)$ of degree $k/\sqrt{K}=kL$. As a result,  (\ref{interesting}) is zero  when $\ell\geq kL$. For $K=+1/L^2$ we therefore  restrict $\ell < kL$.  

Notice that, for  $K=-1/L^2$, $kL$ is not necessarily an integer, so the reasoning above cannot be applied in this case and $\ell$ is in general unbounded.

\subsection{Orthogonality and normalization}

We discuss now  the normalization factor for the field modes. We follow the logic of \cite{BI} (in particular, see its section II.D). 
Before we argued that the radial dependence of the field modes, subject to regularity conditions at the origin, satisfies
\bea
 \Pi_{K, k\ell}(\chi)= A_{k \ell} \sin^{\ell}(\sqrt{K}\chi) \left[\frac{d}{d\cos(\sqrt{K}\chi)} \right]^{\ell+1}\cos(k\chi)\, . \label{pis}
\eea
The full space-like modes $\mathcal Y_{K,\vec k}(\vec x) =  \Pi_{K, k\ell}(\chi)Y_{\ell m}(\theta,\phi)$ are required to be normalized as (\ref{ortho1}). Using the familiar orthogonality condition for the spherical harmonics,  we can partially evaluate this integral:
\bea
 \int_{\Sigma} \mathcal Y_{K,\vec k}(\vec x)  \bar{\mathcal Y}_{K,\vec k'}(\vec x)\,\sqrt{h}\, \rmd^3x\,=\, \delta_{\ell,\ell'}\delta_{m,m'} \frac{| A_{k \ell}|^2}{K} \int_0^{x_K}  q_{\ell+1}(k_1,x)\, q_{\ell+1}(k_2,x)\,\rmd x 
\eea
where $x_K\equiv \pi L$ for $K=1/L^2$ and $x_K\equiv +\infty$ for $K=-1/L^2$. To alleviate the notation we introduced the function
\bea
 q_{\ell}(k,x)=\sin^{\ell}(\sqrt{K}x) \left[\frac{d}{d\cos(\sqrt{K}x)} \right]^{\ell}\cos(kx) \label{defq}
\eea
This function satisfies the following identities, 
\bea
q_{\ell+1}(k,x) & = &  -\frac{1}{\sqrt{K}} \frac{dq_{\ell}(k,x)}{dx} + \ell \cot(\sqrt{K}x)q_{\ell}(k,x)\\
 0 & = &  \left[\partial_x^2+N^2-\frac{K\ell(\ell+1)}{\sin^2(\sqrt{K}x)} \right] q_{\ell}(k,x)\\
(k^2-K(\ell-1)^2)q_{\ell-1}(k,x)  & = &  \sqrt{K}\frac{d q_{\ell}(k,x)}{dx}+ K(\ell-1) \cot(\sqrt{K}x)q_{\ell}(k,x)
\eea
The first identity is easy to check from (\ref{defq}). The second identity is a consequence of the differential equation that satisfies $ \Pi_{K, k\ell}(\chi)$:
\bea
\left[\partial_{\chi}^2+2\sqrt{K}\cot(\sqrt{K}\chi) \partial_{\chi} -\frac{K \ell(\ell+1)}{\sin^2(\sqrt{K}\chi)} +k^2-K\right]\frac{q_{\ell+1}(k,\chi)}{\sin(\sqrt{K}\chi)}=0
\eea
The third identity can be obtained easily from the first two. Using these functional relations we can now compute  the full integral of interest:
\bea
\int_0^{x_K} q_{\ell+1}(k_1,x) q_{\ell+1}(k_2,x) \rmd x  &= &  \int_0^{x_K} q_{\ell+1}(k_1,x) \left[-\frac{1}{\sqrt{K}} q'_{\ell}(k_2,x)+\ell \cot(\sqrt{K}x) q_{\ell}(k_2,x)\right] \rmd x   \nonumber \\
 & = & \int_0^{x_K} q_{\ell}(k_2,x) \left[\frac{1}{\sqrt{K}} q'_{\ell+1}(k_1,x)+\ell \cot(\sqrt{K}x) q_{\ell+1}(k_1,x)\right] \rmd x \nonumber \\
 && -\frac{1}{\sqrt{K}}\left. q_\ell q_{\ell+1}\right|_{0}^{x_K} \label{auxiliar}
\eea 
To proceed with the calculation we will distinguish the two cases.\vskip0.2cm

1) Let $K=1/L^2$. In this case, it easy to find from the definition that $q_{\ell}(k,0)=q_{\ell}(k,\pi L)=0$ for odd $\ell$. Therefore the boundary term above vanishes for all $\ell$. Using one of the identities introduced above, we get
\bea
\int_0^{x_K} q_{\ell+1}(k_1,x) q_{\ell+1}(k_2,x) \rmd x  &= &  \left(\frac{k_1^2}{K}-\ell^2\right) \int_0^{x_K} q_{\ell}(k_1,x) q_{\ell}(k_2,x) \rmd x
\eea
Proceeding recursively:
\bea
\int_0^{x_K} q_{\ell+1}(k_1,x) q_{\ell+1}(k_2,x) \rmd x  &= &\frac{k_1^2}{K} \left(\frac{k_1^2}{K}-1^2\right)\dots \left(\frac{k_1^2}{K}-\ell^2\right) \int_0^{x_K} q_{0}(k_1,x) q_{0}(k_2,x) \rmd x \quad \quad \quad \text{} \label{aux5}
\eea
The last integral is easy to evaluate:
\bea
 \int_0^{x_K} q_{0}(k_1,x) q_{0}(k_2,x) \rmd x = \int_0^{x_K} \cos(k_1 x)\cos(k_2 x) \rmd x =\frac{\pi}{2} L \delta_{k_1,k_2}
\eea
Collecting all these results, and demanding $ \int_{\Sigma}d^3x\sqrt{h}  \mathcal Y_{K,\vec k}(\vec x)  \bar{\mathcal Y}_{K,\vec k'}(\vec x)=L\delta_{k,k'}\delta_{\ell,\ell'}\delta_{m,m'}$, we end up with 
\bea
A_{k \ell}=\frac{e^{i\phi_{k\ell}} \sqrt{K}}{\sqrt{\frac{\pi}{2}  \Pi_{i=0}^{\ell} [k^2/K - i^2] }  } 
\eea
where $\phi_{k\ell}$ is an arbitrary phase. Note that the denominator never vanishes because   $\ell < k L$. Note also that (\ref{pis}) is well-defined for $k=0$ with this normalization factor for any $\ell$. Taking into account that $k^2/K - i^2=(k/\sqrt{K}+i)(k/\sqrt{K}-i)$, this expression can be further written as
\bea
A_{k \ell}=e^{i\phi_{k\ell}}\frac{\sqrt{2 K^{3/2} \Gamma(k/\sqrt{K}-\ell)}}{\sqrt{\pi \, k\,  \Gamma(k/\sqrt{K}+\ell+1)}    } \label{prefactor}
\eea
where we used $\Gamma(k/\sqrt{K}+1)/\Gamma(k/\sqrt{K})=k/\sqrt{K}$.\vskip0.2cm

2) Let $K=-1/L^2$.  In this case $q_{\ell}(k,0)=0$ for odd $\ell$ but  $q_{\ell}(k,\infty)$ is oscillatory. To deal with this case, we  work with an integrated version of (\ref{auxiliar}):
\bea
\int_{-\infty}^{+\infty} \!\!\!\rmd k_2 \int_0^{x_K}\!\! q_{\ell+1}(k_1,x) q_{\ell+1}(k_2,x) dx & = & \int_{-\infty}^{+\infty} \!\!\! \rmd k_2  \int_0^{x_K}\!\! q_{\ell}(k_2,x) \left[\frac{q'_{\ell+1}(k_1,x)}{\sqrt{K}} +\ell \cot(\sqrt{K}x) q_{\ell+1}(k_1,x)\right] \rmd x \nonumber\\
&  & -\lim_{x\rightarrow \infty} \int_{-\infty}^{+\infty}\!\! \rmd k_2  \frac{1}{\sqrt{K}}\, q_\ell(k_2,x)\, q_{\ell+1}(k_1,x) \nonumber
\eea
It is not difficult to see that, when $x\to \infty$, the function $q_{\ell}(k,x)$ asymptotically equals $p_{\ell}(k,x)=f_1(k,\ell)\cos(kx)+f_2(k,\ell)\sin(kx)$ for some functions $f_1$, $f_2$. The product $p_{\ell}(k_2,x)p_{\ell+1}(k_1,x)$ will be a linear combination of $\cos((k_1\pm k_2) x)$, $\sin((k_1-k_2) x)$. Subtracting and summing this function inside the second integral above, and applying the Riemann-Lebesgue lemma, this integral  is zero. Doing now exactly the same calculation as in the $K=+1/L^2$ case,   using the following integral representation of the Dirac delta
\footnote{$\delta(k_1+k_2)=0$ because $k_1, k_2\geq 0$, and in the case $k_1=k_2=0$ the prefactor $k_1^2$ in (\ref{aux5}) makes it vanish}
\bea
 \int_0^{\infty}\!\! q_{0}(k_1,x) q_{0}(k_2,x) \rmd x  & = &  \int_0^{\infty} \cos(k_1 x)\cos(k_2 x) \rmd x = \frac{1}{2}\int_{-\infty}^{\infty} \cos(k_1 x)\cos(k_2 x) \rmd x \nonumber\\
 & =& \frac{1}{4}\int_{-\infty}^{\infty} [\cos((k_1-k_2) x)+\cos((k_1+k_2) x)] \rmd x \nonumber\\
 &  = & \frac{2\pi}{4}( \delta(k_1-k_2)+ \delta(k_1+k_2)) =\frac{\pi}{2}  \delta(k_1-k_2)
\eea
 and demanding $ \int_{\Sigma} \rmd^3x \,\sqrt{h}\, \mathcal Y_{K,\vec k}(\vec x)  \bar{\mathcal Y}_{K,\vec k'}(\vec x)=\delta(k_1-k_2)\delta_{\ell,\ell'}\delta_{m,m'}$, we end up with the same coefficient (\ref{prefactor}).

\subsection{Addition Theorem}

Using (\ref{interesting}) with (\ref{pis}) and (\ref{prefactor}) we can write
\bea
 \Pi_{K, k\ell}(\chi)=i  e^{i\phi_{k\ell}} \sqrt{k}K^{1/4}\sqrt{\frac{\Gamma(k/\sqrt{K}+\ell+1)}{\Gamma(k/\sqrt{K}-\ell)}} \frac{P_{-\frac{1}{2}+ \frac{k}{\sqrt{K}}}^{-(\ell+\frac{1}{2})}(\cos(\sqrt{K}x))}{\sqrt{\sin(\sqrt{K}x)}}
\eea
From the relation with  the Gegenbauer function [see eqn 8.936(1) in \cite{GR}]
\bea
C^{\ell+1}_{k-(\ell+1)}(\cos(\sqrt{K}x)) =  \frac{P_{-\frac{1}{2}+ \frac{k}{\sqrt{K}}}^{-(\ell+\frac{1}{2})}(\cos(\sqrt{K}x))}{\sin^{1/2+\ell}(\sqrt{K}x)}  \frac{(-1)^{-1/4-\ell/2}}{2^{-\ell-1/2}} \frac{\Gamma(\ell+1+k/\sqrt{K})\Gamma(\ell+3/2)}{\Gamma(2\ell+2)\Gamma(k/\sqrt{K}-\ell)} \nonumber
\eea  
we can infer
\bea
 \Pi_{K, k\ell}(\chi)=  e^{i\phi_{k\ell}'} \sqrt{k}K^{1/4} \sin^{\ell}(\sqrt{K}x) C^{\ell+1}_{k-(\ell+1)}(\cos(\sqrt{K}x))  \sqrt{\frac{ \Gamma(k/\sqrt{K}-\ell)}{\Gamma(k/\sqrt{K}+\ell+1)}}\frac{2^{\ell+1/2} }{\sqrt{\pi}}\Gamma(\ell+1) \nonumber
\eea 
where we used the identity $\Gamma(2\ell+2)/\Gamma(\ell+3/2)=\frac{2^{2(\ell+1)-1}}{\sqrt{\pi}}\Gamma(\ell+1)$ [see eqn 8.335(1) in \cite{GR}], and introduced a new phase $\phi_{k\ell}':=\phi_{k\ell}+\frac{5\pi}{4}+\frac{\pi \ell}{2}$. 

Let us compute  the addition formula:
\bea
\sum_{\ell m} \mathcal Y_{K,\vec k}(\vec x) \bar{\mathcal Y}_{K,\vec k}(\vec y) &  = &  \sum_{\ell m}  \Pi_{K, k\ell}(x)  \bar\Pi_{K, k\ell}(y) |Y_{\ell m}(\theta,\phi)|^2 =   \sum_{\ell=0 }^{\infty}  \Pi_{K, k\ell}(x)  \bar\Pi_{K, k\ell}(y) \frac{2\ell+1}{4\pi} \nonumber \\
& = & \frac{k K^{1/2}}{2\pi^2}  \sum_{\ell=0}^{\infty}\sin^{\ell}(\sqrt{K}x) C^{\ell+1}_{k-(\ell+1)}(\cos(\sqrt{K}x)) \sin^{\ell}(\sqrt{K}y) C^{\ell+1}_{k-(\ell+1)}(\cos(\sqrt{K}y))  \nonumber \\
&& \quad \quad \quad \quad \quad \times \frac{ 2^{2\ell}\Gamma(k/\sqrt{K}-\ell) \Gamma(\ell+1)^2}{\Gamma(k/\sqrt{K}+\ell+1)}(2\ell+1)  \nonumber
\eea
where in the second equality we used the addition theorem for the spherical harmonics. Note that, when $K=1/L^2$, the sum is finite as the summand is non-zero only when $\ell=0,\dots, kL-1$.
Using now the addition theorem for Gegenbauer polynomials [see eqn 8.934(3) in \cite{GR}; use the identity $C^{1/2}_{\ell}(\cos 0)=P_{\ell}(1)=1$ from eqn 8.936(3) in \cite{GR}] we obtain
\bea
\sum_{\ell m} \mathcal Y_{K,\vec k}(\vec x)\bar{ \mathcal Y}_{K,\vec k}(\vec y)   =  \frac{k K^{1/2}}{2\pi^2}  C^{1}_{k-1}(\cos(\sqrt{K}(x-y)))
\eea
Using again the relation between   Gegenbauer and Legendre polynomials [eqn 8.936(1) in \cite{GR}] we find $C^{1}_{k-1}(\cos(\sqrt{K}(x-y))) = \frac{k}{\sqrt{K}}\sqrt{\frac{\pi}{2}}\frac{1}{(-4)^{-1/4}}  \frac{1}{\sqrt{\sin(\sqrt{K}(x-y))}}P_{-\frac{1}{2}+ \frac{k}{\sqrt{K}}}^{-\frac{1}{2}}(\cos(\sqrt{K}(x-y)))$. We also  have (see eqn 8.754(3) in \cite{GR}), $P^{-1/2}_{\nu}(\cos(\sqrt{K}x))=\sqrt{\frac{2}{\pi \sin(\sqrt{K}x)}}\frac{\sqrt{K}}{k} \sin(k x)$, so we finally get
\bea
\sum_{\ell m} \mathcal Y_{K,\vec k}(\vec x) \bar{\mathcal Y}_{K,\vec k}(\vec y)   = \frac{k K^{1/2}}{2\pi^2}   \frac{\sin(k (x-y))}{\sin(\sqrt{K}(x-y))}
\eea

%*******

\end{document}